\documentclass[amsmath,prb,aps,twocolumn,showpacs,10pt,citeautoscript]{revtex4-1}

\usepackage{bm}%
\usepackage{graphicx}%
\usepackage{dcolumn}%
\usepackage[breaklinks=true,colorlinks=true,linkcolor=blue,urlcolor=blue,citecolor=blue]{hyperref}

\begin{document}

\title{Thermal and Thermoelectric Properties of Graphene}
\author{Yong \surname{Xu}$^{1,2,3}$}
\email{yongxu@stanford.edu}
\author{Zuanyi \surname{Li}$^{4,5}$}
\email{zuanyili@stanford.edu}
\author{Wenhui \surname{Duan}$^{1,2,6}$}
\email{dwh@phys.tsinghua.edu.cn}
\affiliation{$^1$Department of Physics and State Key Laboratory of Low-Dimensional
Quantum Physics, Tsinghua University, Beijing 100084, People's Republic of China \\
$^2$Institue for Advanced Study, Tsinghua University,Beijing 100084, People's Republic of China \\
$^3$Department of Physics, McCullough Building, Stanford University, Stanford, CA 94305-4045, USA \\
$^4$Department of Physics, University of Illinois at Urbana-Champaign, Urbana, IL 61801, USA \\
$^5$Department of Electrical Engineering, Stanford University, Stanford, CA 94305, USA \\
$^6$Collaborative Innovation Center of Quantum Matter, Tsinghua University, Beijing 100084, People's Republic of China }

\normalsize{}

%\date{}%

\begin{abstract}
The subject of thermal transport at the mesoscopic scale and in low-dimensional systems is interesting for both fundamental research and practical applications. As the first example of truly two-dimensional materials, graphene has exceptionally high thermal conductivity, and thus provides an ideal platform for the research. Here we review recent studies on thermal and thermoelectric properties of graphene, with an emphasis on experimental progresses. A general physical picture based on the Landauer transport formalism is introduced to understand underlying mechanisms. We show that the superior thermal conductivity of graphene is contributed not only by large ballistic thermal conductance but also by very long phonon mean free path (MFP). The long phonon MFP, explained by the low-dimensional nature and high sample purity of graphene, results in important isotope effects and size effects on thermal conduction. In terms of various scattering mechanisms in graphene, several approaches are suggested to control thermal conductivity. Among them, introducing rough boundaries and weakly-coupled interfaces are promising ways to suppress thermal conduction effectively. We also discuss the Seebeck effect of graphene. Graphene itself might not be a good thermoelectric material. However, the concepts developed by graphene research might be applied to improve thermoelectric performance of other materials.
\end{abstract}

\maketitle

\section{Introduction}

The Nobel Prize in Physics in 2010 was awarded to Andre Geim and Konstantin Novoselov for their groundbreaking experiments regarding graphene. Graphene is a monolayer of carbon atoms arranged in a regular hexagonal lattice, representing the thinnest material that nature can provide. The research interest of graphene has been growing explosively in the recent years,\cite{Geim-07NM,Castro-09RMP,Sarma-11RMP,Balandin-11NM} as graphene is of great importance to both fundamental research and practical applications. Graphene is the first truly two-dimensional (2D) material and provide an ideal playground to study low-dimensional physics. Meanwhile the existence of massless Dirac Fermions in graphene enables a new paradigm of ``relativistic'' condensed-matter physics.\cite{Geim-07NM,Castro-09RMP,Sarma-11RMP} This unique material supports exceptional properties that are useful for optics, electronics, magnetics, etc.\cite{Geim-07NM,Castro-09RMP,Sarma-11RMP,Balandin-11NM}

Early-stage graphene research intensively focused on electronic properties.\cite{Geim-07NM,Castro-09RMP,Sarma-11RMP} In comparison, the study of thermal properties begins later, but soon becomes an active field\cite{Balandin-11NM,Balandin-10FNCN,Balandin-11AME,Nika-12JPCM,Sadeghi-12SSC,Pop-12MRSB,Balandin-12MT,Shahil-12SSC,ShiL-12NMTE} after the first measurements of thermal conductivity of graphene by Balandin et al. in 2008.\cite{Balandin-08NaL} Graphene offers new opportunities to the development of thermal and thermoelectrics. Unusual thermal transport phenomena and physics emerge in this unique 2D system. Graphene has been experimentally shown to have superior thermal conductivity.\cite{Balandin-11NM} Furthermore, previous theoretical works predict that thermal conductivity gets divergent with increasing transport length in low-dimensional systems.\cite{lepri03pr,basile06prl,casher71jmp,lippi00jsp,dhar01prl,yang02prl,narayan02prl,saito10prl} In this context, it becomes fundamentally important to study the size-dependent behaviors and to find key factors that drive ballistic-diffusive transition. Moreover, it is known that quantum effects become more important in low-dimensional systems than three-dimensional (3D) systems. The 2D graphene sheet or quasi one-dimensional (1D) graphene nanoribbons (GNRs)\cite{HuangB-09FPC,BaiJW-10MSER,lan2009edge,jiang2010isotopic,LiZY-08PRL,LiZY-10JNN} may support prominent quantum effects in thermal transport, like weak localization caused by quantum interference. Last, but not least, the low dimension of graphene systems could significantly affect the scattering strength of various scattering processes,\cite{nika09prb,xu2014} leading to unusual thermal transport behaviors. All these are issues of fundamental importance. The corresponding research could help us to find approaches to effectively control thermal conduction. As applications, thermal conduction of materials can be improved for solving the serious heat dissipation and breakdown in ever-smaller electronic devices,\cite{Pop-06PIEEE,Pop-10NaR,PangL-10JEM,PangL-12JPD,PangL-12TED,LiaoA-10PRB,Tsai-11APL} or suppressed for realizing thermal insulation in high-power engines and also for enhanced thermoelectric efficiency.\cite{Dresselhaus-07AM,Snyder-08NM}

We review thermal and thermoelectric properties of graphene, motivated by the exotic low-dimensional phenomena and physics as well as the great potential for applications.\cite{Subrina-09EDL,YanZ-12NC,Shahil-12NaL,Goyal-12APL,Goli-14JPS} We will summarize the most recent experimental progresses, provide a general physical picture on underlying mechanisms, and further suggest possible ways to tune thermal and thermoelectric properties of graphene as well as other materials.

\section{Basics of Thermal Transport}

Various theoretical approaches have been used to study thermal transport in graphene, including molecular dynamics (MD),\cite{evans2010thermal,OngZY-11PRB,HuJN-11APL,yang12apl,zhang2011thermal} Boltzmann transport equation (BTE),\cite{nika09prb,Lindsay-10PRB2,Lindsay-10PRB3,Lindsay-11PRB,Aksamija-11APL,Aksamija-12PRB} non-equilibrium Green's function (NEGF),\cite{XuY-09APL,XuY-10PRB,jiang11apl,lu12apl,yang12apl,Serov-13APL} and Landauer approach.\cite{Saito-07PRB,XuY-09APL,XuY-10PRB} Herein we mainly introduce the Landauer approach, which is a widely used theoretical tool for mesoscopic transport.\cite{Datta97} The approach can be generally applied to investigate ballistic-diffusive transport for systems from 1D to 3D.\cite{XuY-08PRB} Importantly, it is conceptually simple, which enables us to build a clear and general physical picture to understand rich thermal transport phenomena in graphene.

In the Landauer transport formalism, the lattice thermal conductance $K_l$ is written as
\begin{align}\label{K_l}
K_{l} = \frac{k_{\rm B}^2 T}{h}  \int_{0}^{\infty} dx \frac{x^2 e^x}{(e^x -1)^2}
\overline{\mathcal T}_{p} (x),
\end{align}
where $k_{\rm B}$ is the Boltzmann constant, $T$ is the absolute temperature, $h$ is the Planck constant, $x = \hbar \omega / (k_{\rm B} T)$, $\hbar$ is the reduced Planck constant, $\omega$ is the phonon frequency, $\overline{\mathcal T}_{p} (x) \equiv \overline{\mathcal T}_{p} (\omega)$ is the phonon transmission function.\cite{XuY-10PRB} $\overline{\mathcal T}_{p} (\omega) = M_{p}(\omega){\mathcal T}_{p}(\omega)$. The distribution of phonon modes $M_{p}(\omega)$ counts the number of phonon transport channels at a given frequency $\omega$, which is equal to the ballistic phonon transmission function. The transmission probability ${\mathcal T}_{p}(\omega)$ is equal to 1 in the ballistic limit and $\lambda (\omega)/L$ in the diffusive limit, where $\lambda (\omega)$ is the phonon mean free path (MFP) for backscattering. When quantum interference effects are neglected, we get a quasi-classical formula
\begin{align}\label{trans_p}
{\mathcal T}_{p}(\omega) = \frac{\lambda (\omega)} {\lambda (\omega) + L},
\end{align}
where $L$ is the transport length.\cite{XuY-08PRB} The formula, which is exact in both ballistic and diffusive limits, can be used to describe the ballistic-diffusive transport.\cite{XuY-10PRB} It is interesting that the Landauer formula reproduces the BTE in the diffusive limit.\cite{XuY-08PRB,jeong10jap} However, one should notice that the MFPs used in the BTE and in the Landauer approach are not exactly the same. The BTE uses a common MFP defined as the average transport distance between two successive scatterings. While the Landauer approach uses the MFP for backscattering, since in the Landauer approach transport quantities are determined by transmission, to which only backscattering is related. If assuming isotropic phonon bands, the ratio of backscattering MFP to common MFP is $2$, $\pi/2$ and $4/3$  for 1D, 2D and 3D systems, respectively.\cite{jeong10jap} Without specification, we will use $\lambda$ to denote the MFP for backscattering, which is the MFP relevant to transport in the Landauer framework.

By assuming $\overline{\mathcal T}_{p} \equiv 1$ in Equation (\ref{K_l}), we get $K_0 = \pi^{2} k_B^2 T/3h = (9.456 \times 10^{-13}~{\rm WK}^{-2})T$. $K_0$, called as thermal conductance quantum, represents the maximum possible value of energy transported per conduction mode. In contrast to electrical conductance quantum, thermal conductance quantum is not a constant but proportional to temperature. The existence of the quantum of thermal conductance was theoretically predicted in 1998~\cite{Rego-98PRL} and  experimentally proved in 2000.\cite{Schwab-00N} Equation (\ref{K_l}) shows that thermal conductance is a weighted integration of phonon transmission function. The weighting factor, ${x^2 e^x/(e^x -1)^2}$, plays an essential role in determining thermal conduction contribution. The factor is equal to one at $x=0$, decreases quickly with increasing $x$, and becomes vanishingly small when $x > 10$. This tells us that thermal conduction are mainly contributed by low frequency phonons, and  phonons with frequency higher than $10 k_{\rm B}T/\hbar$ have negligible contribution to thermal conduction.

In theoretical calculations, $M_{p}(\omega)$ is determined from phonon dispersion by counting transport channels. The ballistic thermal conductance $K_l^{\rm{ball}}$ as a function of temperature is then obtained from Equation (\ref{K_l}). Thermal transport
studies usually use the approximation of constant phonon MFP ($\lambda(\omega) \equiv \lambda$), which does not rely on any assumption about the possible dependence on temperature of the phonon MFP. In this approximation, $K_l$ and $K_l^{\rm ball}$ are related by $K_l = K_l^{\rm{ball}} \lambda/(\lambda + L)$, which gives $K_l^{\rm diff} = K_l^{\rm ball} \lambda / L$ in the diffusive limit ($L\gg\lambda$). The lattice thermal conductivity $\kappa_l$ is defined as $\kappa_l = K_l L/A$, where $A$ is the cross-sectional area. The diffusive $\kappa_l^{\rm diff}$ is thus given by\cite{xu2014,Bae-13NC}
\begin{align}\label{kappa_l}
\kappa_l^{\rm diff} = \lambda K_l^{\rm{ball}} / A.
\end{align}
This is a very useful relation. As $K_l^{\rm{ball}} / A$ is easily obtained by theoretical calculations and the diffusive $\kappa_l^{\rm diff}$ is measured by experiments, the combined information from theory and experiment determines $\lambda$, one of the most important length scales in thermal transport. Based on the estimated $\lambda$, we know that thermal transport is ballistic if $L \ll \lambda$, or diffusive if $L \gg \lambda$, or in the intermediate region otherwise.

Thermal conduction in graphene is mainly contributed by lattice vibrations (i.e., phonons), and the contribution of electrons is expected to be negligible according to theoretical calculations\cite{Saito-07PRB,Watanabe-09PRB}. Indeed, very recent experiments\cite{Fong-13PRX,Yigen-13PRB,Yigen-14NaL} showed electron thermal conductivity is much less than total thermal conductivity ($<$ 1\%) in suspended graphene. We thus focus on discussing the lattice vibration contributed thermal conduction. Hereafter, without specification we simply use $K$ and $\kappa$ without the subscript ``$l$'' to denote thermal conductance and thermal conductivity of lattice vibrations.

\section{Experimental Methods}
\label{expt.method}

\begin{figure*} [tbp]
\centering
\includegraphics[width=0.8\textwidth]{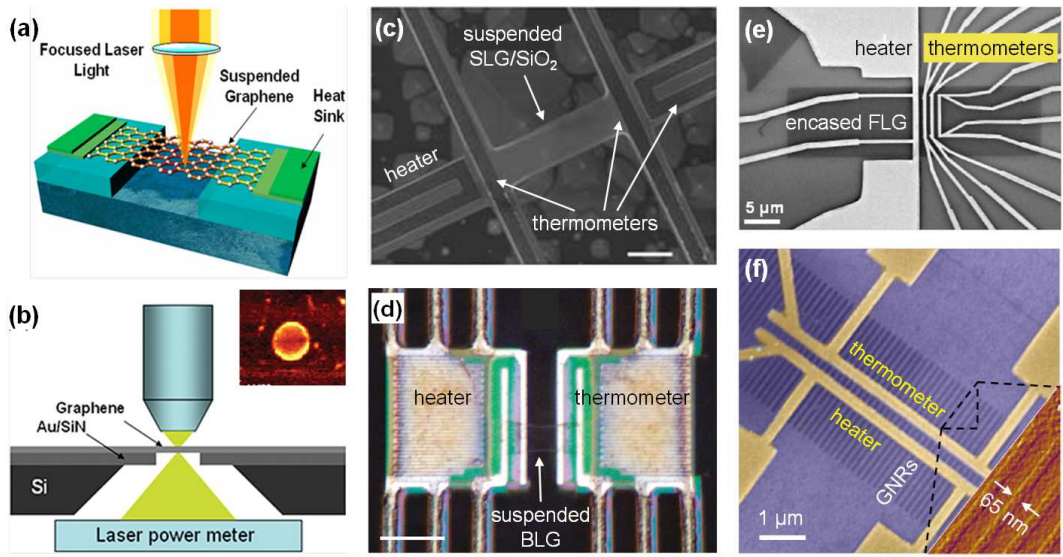}
\caption{\label{fig:thermometry} a) Schematic of the optothermal Raman thermometry set-up, where a graphene strip is suspended over a trench and heated up by a focused laser light. Reproduced with permission.\cite{Ghosh-09NJP} Copyright 2009, IOP. b) Schematic of the Raman thermometry set-up with addition of a laser power meter to measure optical transmittance. Inset is the Raman G peak map of graphene suspended over a circular hole. Reproduced with permission.\cite{CaiWW-10NaL} Copyright 2010, ACS. c) Scanning electron microscopy (SEM) image of micro-resistance thermometry device with SLG supported on a suspended SiO$_2$ membrane between thermometers. Scale bar is 3 $\mu$m. Reproduced with permission.\cite{Seol-10S} Copyright 2010, AAAS. d) Optical image of bilayer graphene (BLG) suspended over two thermometer pads. Scale bar is 10 $\mu$m. Reproduced with permission.\cite{Pettes-11NaL} Copyright 2011, ACS. e) SEM image of a SiO$_2$/Si-supported micro-resistance thermometry device to measure encased few-layer graphene (FLG). Reproduced with permission.\cite{Jang-10NaL} Copyright 2010, ACS. f) False-colored SEM image of a GNR array on SiO$_2$/Si with micro-resistance thermometers. Inset is a zoom-in atomic force microscopy (AFM) image of GNRs. Reproduced with permission.\cite{Bae-13NC} Copyright 2013, NPG.}
\end{figure*}

Measuring nanoscale thermal transport is quite challenging due to high requirements of sample fabrication and temperature sensing.\cite{Cahill-12JHT,Brites-12NaS} So far, methods used to probe thermal conduction in graphene include optothermal Raman thermometry,\cite{Balandin-08NaL,Ghosh-08APL,Ghosh-09NJP,Ghosh-10NM,CaiWW-10NaL,LeeJU-11PRB,ChenSS-11ACSN,ChenSS-12NM,ChenSS-12NaT,Faugeras-10ACSN,Vlassiouk-11NaT,Ermakov-13NaT} thermoreflectance technique,\cite{Mak-10APL,Koh-10NaL,Hopkins-12NaL,ZhangCW-13Carbon} 3$\omega$ method,\cite{ChenZ-09APL} micro-resistance thermometry,\cite{Seol-10S,Seol-10JHT,WangZQ-11NaL,Sadeghi-13PNAS,Pettes-11NaL,WangZQ-11NaL,Jang-10NaL,Bae-13NC,Jang-13APL} electrical self-heating method,\cite{Dorgan-13NaL,XieHQ-13APL} and scanning thermal microscopy (SThM).\cite{Yu-11APL,Pumarol-12NaL,Menges-13PRL} Here we mainly discuss and compare two techniques widely used to measure in-plane thermal conductivity of graphene, i.e., optothermal Raman thermometry and micro-resistance thermometry.

The optothermal Raman thermometry technique was developed by Balandin et al.\cite{Balandin-08NaL} to measure \emph{suspended}, \emph{micrometer} scale ($> 2~\mu$m) graphene. A laser light was focused at the center of the suspended graphene flake to generate a heating power $P_{\rm H}$ and raise temperature locally (\textbf{Figure \ref{fig:thermometry}a and \ref{fig:thermometry}b}). Meanwhile, the Raman spectrum of graphene was recorded and the temperature rise $\Delta T$ could be monitored by calibrating it with Raman G peak position,\cite{Balandin-08NaL,Ghosh-08APL,Ghosh-09NJP,Ghosh-10NM,CaiWW-10NaL} 2D peak position\cite{LeeJU-11PRB,ChenSS-11ACSN,ChenSS-12NM,ChenSS-12NaT} (only for monolayer), or Stokes/anti-Stokes ratio.\cite{Faugeras-10ACSN,Vlassiouk-11NaT} By knowing the correlation between $P_{\rm H}$ and $\Delta T$, as well as the geometry size of suspended graphene, its in-plane thermal conductivity $\kappa$ can be extracted through the solution of the heat diffusion equation. The early experiments\cite{Balandin-08NaL,Ghosh-08APL,Ghosh-09NJP,Ghosh-10NM} were carried out on graphene strips suspended over a trench (Figure \ref{fig:thermometry}a). This was modified in subsequent experiments\cite{CaiWW-10NaL,Faugeras-10ACSN,LeeJU-11PRB,ChenSS-11ACSN,ChenSS-12NM,ChenSS-12NaT} by adopting circular holes with graphene over them (Figure \ref{fig:thermometry}b), which matches with the radial symmetry of the laser spot and allowed for an analytic solution of the temperature distribution.\cite{Sadeghi-12SSC}
%where low geometry symmetry might introduce extra uncertainty in temperature estimations.

A major source of uncertainty in different works using the Raman thermometry technique is determining the laser power absorbed by graphene, that is, determining optical absorbance of graphene. Balandin et al.\cite{Balandin-08NaL} and Ghosh et al.\cite{Ghosh-08APL,Ghosh-09NJP,Ghosh-10NM} evaluated this number by comparing the integrated Raman G peak intensity of graphene with that of highly oriented pyrolytic graphite (HOPG), leading to $\sim$9\% if converted to optical absorbance for exfoliated single-layer graphene (SLG).\cite{Sadeghi-12SSC} The measured value of $\sim$9\% considered two passes of light (down and reflected back) and resonance absorption effects due to close proximity of graphene to the substrate, and corresponded to 488-nm wavelength where absorption is higher than the 2.3\% long-wavelength limit.\cite{Nair-08S} Faugeras et al.,\cite{Faugeras-10ACSN} Lee et al.,\cite{LeeJU-11PRB} and Vlassiouk et al.\cite{Vlassiouk-11NaT} did not measure the optical absorbance under the conditions of their experiments and assumed a value of 2.3\% for exfoliated SLG based on a separate optical transmission measurement.\cite{Nair-08S} Cai et al.\cite{CaiWW-10NaL} and Chen et al.\cite{ChenSS-11ACSN} obtained values of 3.3$\pm$1.1\% and 3.4$\pm$0.7\% for CVD SLG by directly measuring the optical transmittance via addition of a power meter under the suspended portion of graphene (Figure \ref{fig:thermometry}b). The used optical absorbance is very important because it would proportionally change extracted $\kappa$. We note that both theory\cite{YangL-09PRL} and experiments\cite{KimKS-09N,Kravets-10PRB,Mak-11PRL} showed an increase of optical absorbance in graphene with decreasing laser wavelength due to many-body effect, and the values of Cai et al.\cite{CaiWW-10NaL} and Chen et al.\cite{ChenSS-11ACSN} are consistent with those experimental results. To obtain reliable $\kappa$, it is thus necessary to measure optical absorbance under used laser wavelength and specific experimental conditions.

Another uncertainty source is the calibration of temperature with features of Raman spectrum. It is known that strains and impurities in graphene can affect the Raman peak positions and their temperature dependence,\cite{Ferrari-11NN} which greatly limits the temperature sensitivity of the Raman thermometry technique. In addition, heat loss from graphene to the surrounding air was neglected in most experiments, but Chen et al.\cite{ChenSS-11ACSN} found that for a large diameter (9.7 $\mu$m) graphene flake, the $\kappa$ obtained in air could be overestimated by 14$-$40\% compared with the value obtained in vacuum. This implies measurable errors in previous experiments, even though the influence might be weaker due to smaller sizes of measured graphene. Furthermore, extra uncertainty could come from the difference in $\kappa$ between suspended and supported portions of graphene, as well as thermal boundary resistance between graphene and supporting substrates.\cite{CaiWW-10NaL} Overall, the Raman thermometry technique provides an efficient way to measure $\kappa$ of suspended graphene with benefits of relatively easy sample fabrication, reduced graphene contamination, and simple data analysis, but it inevitably has limitations: (i) relatively large uncertainty (up to 40\%);\cite{Balandin-11NM} (ii) difficulty to probe the low temperature regime due to significant heating in graphene by laser; (iii) inability to be applied to nanometer scale or supported graphene, where edge and interface will take effect.

The micro-resistance thermometry technique is a steady-state method to directly probe heat flows in materials.\cite{Cahill-12JHT} It is able to measure both suspended and \emph{supported} graphene, as well as at the \emph{nanometer} scale and in \emph{low temperature} range, with high resolution of temperature by employing electrical resistance as thermometers. This technique can be further divided into two kinds: suspended bridge platform and fully substrate-supported platform. The former was first developed by Shi et al.\cite{ShiL-03JHT,Moore-11MST} to measure thermal conductivity of 1D nanostructures, and it has been widely used for nanotubes\cite{Kim-01PRL,YuC-05NaL,ChangCW-06PRL,ChangCW-08PRL,Pettes-09AFM,YangJK-11Small} and nanowires.\cite{LiDY-03APL,Hochbaum-08N,Boukai-08N,ChenRK-08PRL,Lim-12NaL} By using this platform, graphene can be either supported by a suspended SiO$_2$/SiN$_{\rm x}$ membrane connecting two thermometers\cite{Seol-10S,Seol-10JHT,WangZQ-11NaL,Sadeghi-13PNAS} (\textbf{Figure \ref{fig:thermometry}c}) or fully suspended over two thermometer pads\cite{Pettes-11NaL,WangZQ-11NaL} (\textbf{Figure \ref{fig:thermometry}d}), enabling measurements of both suspended and supported graphene. The fully substrate-supported platform was developed by Jang et al.\cite{Jang-10NaL} and Bae et al.,\cite{Bae-13NC} where at least two thermometers were patterned on Si/SiO$_2$-supported graphene (\textbf{Figure \ref{fig:thermometry}e}) or GNRs (\textbf{Figure \ref{fig:thermometry}f}). In both platforms, one thermometer serves as the heater to generate heating power $P_{\rm H}$ and a temperature gradient across graphene by electrical heating, meanwhile all thermometers (including the heater) monitor temperature changes $\Delta T$ in terms of their electrical resistance changes. Then, the thermal conductance/conductivity of measured materials can be extracted in a simple analytic way by solving its equivalent thermal resistance circuit for the suspended bridge platform,\cite{ShiL-03JHT,Moore-11MST} while a complicated 3D numerical (finite element) simulation has to be performed for the substrate-supported platform due to significant heat leakage into the substrate.\cite{Jang-10NaL,Bae-13NC}

Although the data extraction of the suspended bridge platform is easier than that of the substrate-supported platform, as a trade-off the sample fabrication is more complicated for the former than the latter. Thus, for materials which are hard to be suspended, such as GNRs, the latter is an advantageous method to be employed. However, the measurable length of interested materials cannot be longer than a few micrometer for the substrate-supported platform, because the temperature drops nearly exponentially away from the heater, leading to undetectable $\Delta T$ if other thermometers are far away. It is worth noting that for the platform of graphene supported by a suspended membrane (Figure \ref{fig:thermometry}c) and the substrate-supported platform (Figure \ref{fig:thermometry}e and \ref{fig:thermometry}f), a control experiment has to be carried out by etching off graphene/GNRs and repeating measurements to calibrate the background heat flow and thermal contact resistance between graphene and thermometers.\cite{Seol-10S,Seol-10JHT,Jang-10NaL,Bae-13NC} This improves the measurement accuracy. Whereas, for the platform of graphene fully suspended (Figure \ref{fig:thermometry}d) such a control experiment cannot be performed, so the thermal contact resistance could be a main source of uncertainty in results. Overall, the micro-resistance thermometry technique has very high resolution of temperature ($<$ 50 mK)\cite{Balandin-11NM,Sadeghi-12SSC} and can cover a wide temperature range. It can probe both suspended and supported graphene, as well as in nanometer scale. However, attention should be paid to micro/nanofabrication, which could introduce contaminations (like residues and defects) and increase the uncertainty.\cite{Pettes-11NaL}

\section{Intrinsic Thermal Conductivity of Graphene}

In this section, we mainly discuss the ``intrinsic" thermal conductivity of SLG based on experimental results and theoretical analysis. Here, by ``intrinsic" we mean isolated, large scale, pristine graphene without suffering impurity, defect, interface, and edge scatterings, so its thermal conductivity is only limited by intrinsic phonon-phonon scattering due to crystal anharmonicity\cite{Balandin-11NM} and electron-phonon scattering. In experiments, suspended, micrometer scale graphene samples have properties close to intrinsic ones. We thus first summarize current experimental observations of $\kappa$ in suspended SLG, then discuss the underlying physical origins of high thermal conductivity in graphene.

\begin{figure*} [tbp]
\centering
\includegraphics[width=0.75\textwidth]{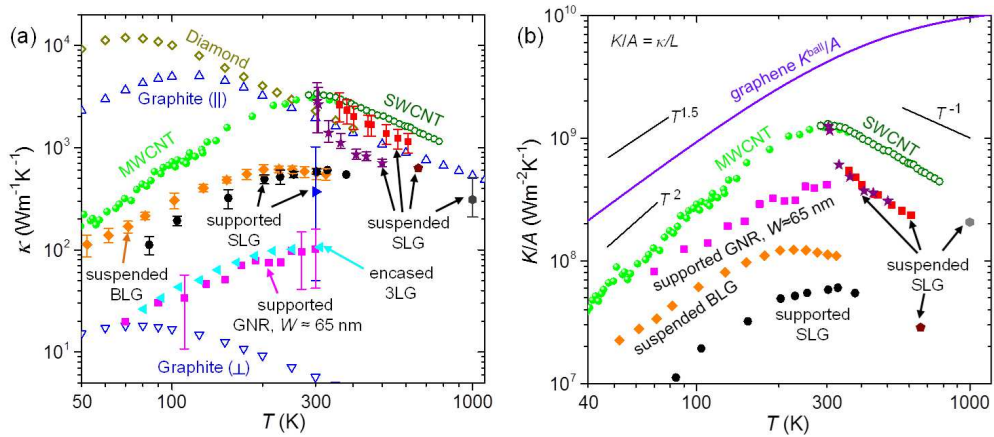}
\caption{\label{fig:kg-T} a) Experimental thermal conductivity $\kappa$ as a function of temperature $T$: representative data for suspended CVD SLG by Chen et al.\cite{ChenSS-11ACSN} (solid red square), suspended exfoliated SLG by Lee et al.\cite{LeeJU-11PRB} (solid purple asterisk) and Faugeras et al.\cite{Faugeras-10ACSN} (solid brown pentagon), suspended SLG by Dorgan et al.\cite{Dorgan-13NaL} (solid grey hexagon), suspended exfoliated BLG by Pettes et al.\cite{Pettes-11NaL} (solid orange diamond), supported exfoliated SLG by Seol et al.\cite{Seol-10S} (solid black circle), supported CVD SLG by Cai et al.\cite{CaiWW-10NaL} (solid blue right-triangle), encased exfoliated 3-layer graphene (3LG) by Jang et al.\cite{Jang-10NaL} (solid cyan left-triangle), supported exfoliated GNR of $W\approx65$ nm by Bae et al.\cite{Bae-13NC} (solid magenta square), type IIa diamond\cite{Ho-72JPCRD} (open gold diamond), graphite in-plane\cite{Ho-72JPCRD} (open blue up-triangle), graphite cross-plane (open blue down-triangle), suspended single-walled CNT (SWCNT) by Pop et al.\cite{Pop-06NaL} (open dark-green circle), and multi-walled CNT (MWCNT) by Kim et al.\cite{Kim-01PRL} (solid light-green circle). b) Thermal conductance per unit cross-sectional area, $K/A=\kappa/L$, converted from thermal conductivity data in a), compared with the theoretical ballistic limit of graphene (solid line), which can be approximated analytically as $K^{\rm ball}/A \approx [1/(4.4\times10^5 T^{1.68})+1/(1.2\times10^{10})]^{-1}$ Wm$^{-2}$K$^{-1}$ over the temperature range 1$-$1000 K.\cite{Bae-13NC} Data in a) whose sample $L$ is unknown or not applicable are not shown in b).}
\end{figure*}

\subsection{Experimental Results}

Using the Raman thermometry technique described above, suspended micro-scale graphene flakes obtained by both exfoliation from graphite\cite{Balandin-08NaL,Ghosh-08APL,Ghosh-09NJP,Ghosh-10NM,LeeJU-11PRB,Faugeras-10ACSN} and CVD growth\cite{CaiWW-10NaL,ChenSS-11ACSN,ChenSS-12NM,ChenSS-12NaT,Vlassiouk-11NaT} have been measured at room temperature and above. Some representative data versus temperature from these studies are shown in \textbf{Figure \ref{fig:kg-T}a}. The obtained in-plane thermal conductivity values of suspended SLG generally fall in the range of $\sim$2000-4000 Wm$^{-1}$K$^{-1}$ at room temperature, and decrease with increasing temperature, reaching about 700-1500 Wm$^{-1}$K$^{-1}$ at $\sim$500 K. The variation of obtained values could be attributed to different choices of graphene optical absorbance (see Section \ref{expt.method}), thermal contact resistance, different sample geometries, sizes, and qualities. For comparison, we also plot the experimental thermal conductivity of diamond,\cite{Ho-72JPCRD} graphite,\cite{Ho-72JPCRD} and carbon nanotubes (CNTs)\cite{Kim-01PRL,Pop-06NaL} in Figure \ref{fig:kg-T}a. It is clear that suspended graphene has thermal conductivity as high as these carbon allotropes near room temperature, even higher than its 3D counterpart, graphite, whose highest record of observed in-plane $\kappa$ in HOPG is $\sim$2000 Wm$^{-1}$K$^{-1}$ at 300 K. The presently available data of graphene based on the Raman thermometry technique only cover the temperature range of $\sim$300-600 K, except one at $\sim$660 K reported by Faugeras et al.\cite{Faugeras-10ACSN} showing $\kappa\approx$ 630 Wm$^{-1}$K$^{-1}$. For higher temperature, Dorgan et al.\cite{Dorgan-13NaL} used the electrical breakdown method for thermal conductivity measurements and found $\kappa\approx$ 310 Wm$^{-1}$K$^{-1}$ at 1000 K for suspended SLG. The overall trend of present graphene data from 300 K to 1000 K shows a steeper temperature dependence than graphite (see Figure \ref{fig:kg-T}a), consistent with the extrapolation of thermal conductivity by Dorgan et al.\cite{Dorgan-13NaL} This behavior could be attributed to stronger second-order three-phonon scattering ($\tau\sim T^{-2}$) in graphene than graphite enabled by the flexural (ZA) phonons of suspended graphene,\cite{Nika-12NaL} similar to the observations in CNTs.\cite{Mingo-05NaL,Pop-06NaL} For temperature below 300 K, the micro-resistance thermometry technique needs to be employed for $\kappa$ measurements. Unfortunately, there is no reliable data for suspended \emph{single}-layer graphene until now. However, data do exist for suspended \emph{few}-layer graphene (FLG),\cite{Pettes-11NaL,WangZQ-11NaL,Jang-13APL} which will be discussed in Section \ref{few-layer}. It is instructive to compare experimental results with the ballistic limit of graphene as a check, so we convert measured $\kappa$ in Figure \ref{fig:kg-T}a to thermal conductance per unit cross-sectional area, $K/A=\kappa/L$, which are re-plotted in \textbf{Figure \ref{fig:kg-T}b} with graphene $K^{\rm ball}/A$ (discussed in the next section). Above room temperature, measured $K/A$ of suspended SLG are more than one order of magnitude lower than $K^{\rm ball}/A$, indicating the diffusive regime. The reason that the value of Faugeras et al.\cite{Faugeras-10ACSN} is much lower than others is because of a much larger $L=22~\mu$m (radius) of their suspended graphene.

\subsection{Ballistic Thermal Conductance}

\begin{figure*}
\includegraphics[width=0.8\linewidth]{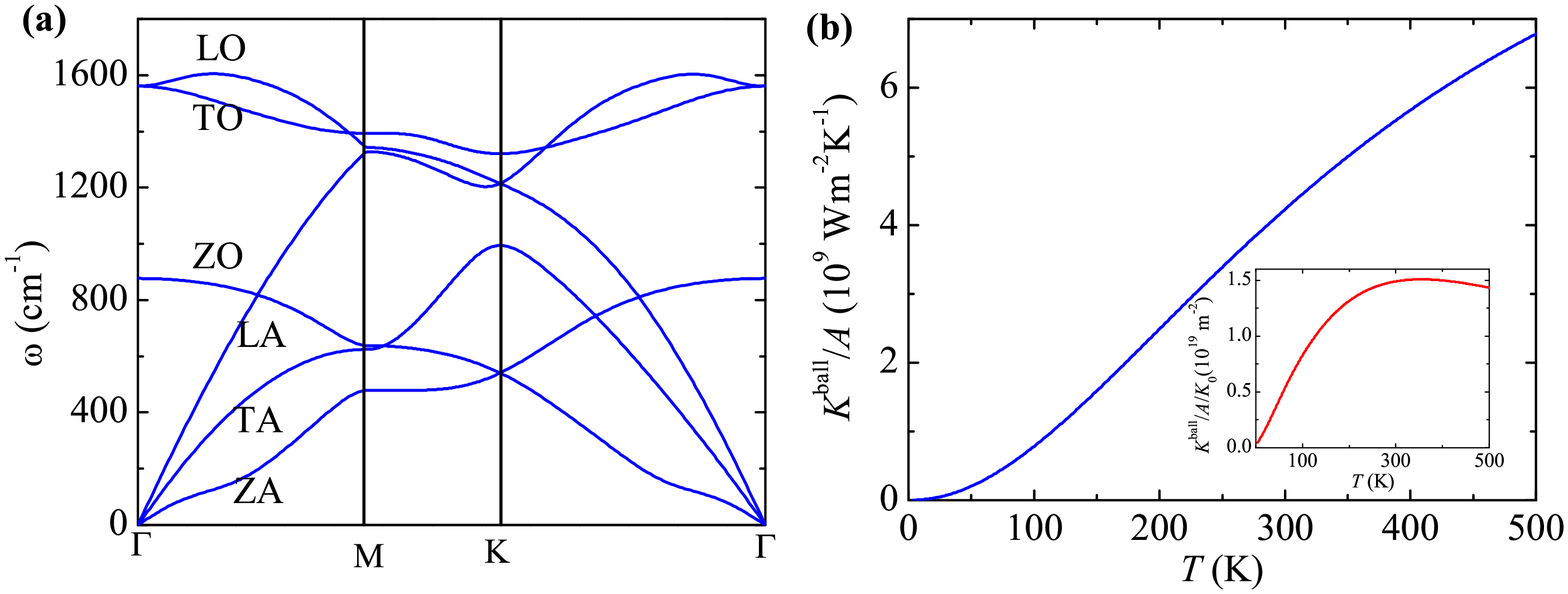}
\caption{\label{phonon} a) Phonon dispersion of graphene computed by density functional theory using the Perdew-Burke-Ernzerhof-type exchange-correlation functional\cite{perdew96prl} in the generalized-gradient approximation as implemented in the VASP code.\cite{kresse96prb} Note that the ZA mode of graphene should have a quadratic dispersion near the $\Gamma$ point, which is not the case here because the rotational symmetry is broken by numerical errors of first principles calculations\cite{bonini2012acoustic}. Such an inaccuracy, however, has minor influence on calculating ballistic thermal conductance according to our tests. b) The ballistic thermal conductance per unit cross-sectional area $K^{\rm{ball}} / A$ of graphene as a function of temperature $T$. The inset shows $K^{\rm{ball}} / A$ scaled by thermal conductance quantum $K_0$.}
\end{figure*}

Graphene has been demonstrated experimentally to possess very high thermal conductivity.\cite{Balandin-11NM,Nika-12JPCM,Sadeghi-12SSC,Pop-12MRSB} What distinguishes graphene from conventional materials in thermal conduction? What is the underlying mechanism to the superior thermal transport ability of graphene?

According to Equation (\ref{kappa_l}), two mechanisms could induce high $\kappa$: (i) large ballistic thermal conductance per unit area $K^{\rm{ball}} / A$  and (ii) long phonon MFP $\lambda$. It is crucial to distinguish the two possibilities, which lead to essentially different strategies to tune thermal conduction in graphene. If graphene has a superiorly large $K^{\rm{ball}} / A$, $\kappa$ would keep large in small samples. While, if $\lambda$ of graphene is extraordinarily long, $\kappa$ would exhibit a strong size dependence, and thermal transport in small samples is ballistic or quasi-ballistic, accompanied by a small $\kappa$ that can be enhanced by increasing $L$. In this context, it becomes important to calculate ballistic thermal conductance.

Let us first look at the phonon dispersion of graphene (see \textbf{Figure \ref{phonon}a}), which was discussed previously, for instance, in Refs. ~\onlinecite{Nika-11PSSB,Cocemasov-13PRB}. Graphene has two carbon atoms in a unit cell, resulting in six phonon bands: three acoustic and three optical bands. These include in-plane transverse acoustic (TA) and optical (TO) modes, in-plane longitudinal acoustic (LA) and optical (LO) modes, and out-of-plane acoustic (ZA) and optical (ZO) modes. For graphene, phonon bands are very dispersive, and the maximal phonon frequency $\omega_{\rm{max}}$ is about 1600 cm$^{-1}$, which is quite high compared to other materials (e.g., $\omega_{\rm{max}} \sim$ 500 cm$^{-1}$ in silicon\cite{yu99fundamentals}). These features, explained by the strong $sp^2$ bonds and the light atomic mass of graphene, are favorable for thermal conduction.

The ballistic thermal conductance of graphene has been discussed previously based on empirical force field.\cite{Mingo-05PRL,Saito-07PRB,Bae-13NC} Here we combine density function theory and Landauer approach to calculate $K^{\rm{ball}} / A$ of graphene as a function of temperature, getting results consistent with previous work.\cite{Mingo-05PRL,Saito-07PRB,Bae-13NC} Herein $A = W \delta$, where $W$ is the width, and $\delta = 0.335$ nm is the the effective thickness selected as the layer separation in graphite. As shown in \textbf{Figure \ref{phonon}b}, $K^{\rm{ball}} / A$ is zero at $T = 0$ K and increases monotonically with temperature. It is as large as $4.2 \times 10^9$ Wm$^{-2}$K$^{-1}$ at $T = 300$ K. The same value has been obtained for  GNRs with zigzag edges.\cite{XuY-09APL}

At this point, it is worthwhile to compare $K^{\rm{ball}} / A$ between graphene and other materials. The room-temperature $K^{\rm{ball}} / A$ is around 1.0 $\times 10^9$ Wm$^{-2}$K$^{-1}$ in silicon,\cite{jeong11jap} smaller than that of graphene but on the same order. It decreases one order of magnitude in Bi$_2$Te$_3$ to around 1.0 $\times 10^8$ Wm$^{-2}$K$^{-1}$.\cite{jeong11jap} Moreover, we compute the ballistic thermal conductance for a newly found 2D topological insulator material, fluorinated stanene (Sn-F), which is a monolayer tin (Sn) film in a honeycomb lattice decorated by fluorine.\cite{XuY-13PRL} We choose to compare $K^{\rm{ball}} / W$ between 2D materials for avoiding selecting a somewhat arbitrary thickness. At room temperature, $K^{\rm{ball}} / W$ of graphene is 1.4 Wm$^{-1}$K$^{-1}$, and that of Sn-F is 0.14 Wm$^{-1}$K$^{-1}$, one order of magnitude smaller.

We suggest a simple way to estimate ballistic thermal conductance of materials, by defining a scaled ballistic thermal conductance $\overline{K}^{\rm{ball}} = K^{\rm{ball}} / (A K_0)$, where $K_0$ is the thermal conductance quantum and is proportional to $T$. This is based on the observation that $K^{\rm{ball}}/A$ is roughly proportional to $T$ in a wide temperature range around room temperature (e.g., between 100 K and 500 K), as evidenced by calculations for graphene (see the inset of Figure~\ref{phonon}b), Sn-F (data not shown), silicon.\cite{jeong11jap} and Bi$_2$Te$_3$.\cite{jeong11jap} In graphene, $\overline{K}^{\rm{ball}}$ is about 10$^{19}$ m$^{-2}$ around room temperature. The value generally decreases when the material compositions change from light to heavy elements. Importantly, the possible decrease is typically equal to or less than one order of magnitude. We get $\overline{K}^{\rm{ball}}$ $\sim$10$^{18}$-10$^{19}$ m$^{-2}$, which depends insensitively on materials type and temperature (around room temperature). This information on the one hand helps us estimate materials ballistic thermal conductance, on the other hand indicates that graphene, if used as a ballistic thermal conductor, is not orders of magnitude better than conventional materials, like silicon.

\subsection{Phonon Mean Free Path and Scattering}

Graphene indeed shows better capability on ballistic thermal conduction than other materials, due to its strong chemical bond and light atomic mass. However, the moderate enhancement in $K^{\rm{ball}}/A$ alone does not explain the superior thermal conductivity of graphene. This leads us to analyze the phonon MFP. We estimate the room-temperature $\lambda$ of graphene using Equation (\ref{kappa_l}) and compare with other materials. For freely suspended graphene, $\kappa$ is typically about 2000-4000 Wm$^{-1}$K$^{-1}$ in experiments,\cite{Balandin-11NM,Nika-12JPCM,Sadeghi-12SSC,Pop-12MRSB} leading to backscattering MFP $\lambda\sim500-1000$ nm (or $\sim300-600$ nm for common MFP).\cite{Bae-13NC} For silicon, the experimental $\kappa$ is 150 Wm$^{-1}$K$^{-1}$, and we obtain $\lambda \sim 150$ nm. For Bi$_2$Te$_3$, experimental $\kappa$ is about 1.5 Wm$^{-1}$K$^{-1}$,\cite{goldsmid1986} giving $\lambda \sim 15$ nm. The results show that the room-temperature $\lambda$ of graphene is about 5 times higher than that of silicon and 50 times than that of Bi$_2$Te$_3$.

Naturally we would ask: why does graphene support a long phonon MFP? To answer this question, let us first look at how phonon MFP or scattering time $\tau$ is determined. The quantities $\lambda$ and $\tau$ are controlled by various scattering processes. According to the nature of scattering, the processes are classified into groups: (i) geometric scattering and (ii) many-body scattering\cite{xu2014}.  The former group includes scattering processes caused by structural imperfections, such as defect scattering, boundary scattering, interface scattering, and isotope scattering. The latter group comprises scattering processes induced by interactions with quasi-particles or collective excitations, like electron-phonon scattering and phonon-phonon scattering. A major difference between these two groups is that the temperature dependence of $\tau$ is usually unimportant in geometric scattering, but becomes significant in many-body scattering. If assuming that scattering processes are independent of each other and each type of scattering contributes $\tau_i$, the total $\tau$ is give by Matthiessen's rule  $1/\tau  = \sum\limits_i {1/\tau _i}$.

Then we come back to the case of graphene. Distinct from conventional materials, graphene has a unique low-dimensional structure, allowing phonons transporting freely within the 2D plane. Moreover, graphene possesses chemical bonds among the strongest in nature, with the $sp^2$ bonds even stronger than the $sp^3$ bonds in diamond.\cite{Pop-12MRSB} These characteristics lead to long phonon MFP in graphene, explained by several mechanisms.\cite{xu2014}

First, the dimensionality of transport systems plays an essential role in thermal conduction. It is known that many-body scattering, like electron-phonon and phonon-phonon scatterings, must satisfy the momentum and energy conservation laws. The satisfaction would be more difficult in systems of lower dimension, because less initial and final states are available for scattering in the phase space~\cite{nika09prb,Nika-09APL}. As a result, the scattering rate decreases considerably in low-dimensional systems. Interestingly, previous works show that the intrinsic $\kappa$ diverges in 1D and 2D systems,\cite{lepri03pr,basile06prl} which questions the validity of Fourier's law. Continuing efforts have been devoted to this subject,\cite{lepri03pr,basile06prl,casher71jmp,lippi00jsp,dhar01prl,yang02prl,narayan02prl,ChangCW-08PRL,saito10prl} but whether Fourier's law works for low-dimensional systems or not remains an open question. Graphene is an ideal playground to study this fundamental problem. The existing experiments find very long but not infinite phonon MFP in graphene. However, to check the divergence of intrinsic $\kappa$, further studies on size effects in large and pure graphene samples are required.

Second, high-quality graphene samples can be realized in experiments, leading to weak defect scattering. It is known that the formation of structural defects (like vacancies, substitutions, and grain boundaries) costs energy to break chemical bonds between adjacent atoms. In graphene, the $sp^2$ bond is very strong, with a bonding energy of 5.9 eV.\cite{Pop-12MRSB} This results in high defect formation energy and thus low defect concentrations in experimental samples.

Third, thermal transport in graphene is insensitive to structural deformation that preserves the $sp^2$ bonding configuration. Although most studies assume that graphene is atomically flat, the strictly 2D crystal could not exist in nature according to the arguments of Landau and Peierls.\cite{Geim-07NM} In fact, ripples develop in graphene to make the system thermodynamically stable.\cite{Fasolino-07NM} Moreover, graphene samples in experiments are often supported by substrates, which could exert strains on samples. The corrugation and strain inevitably scatter phonons and lower thermal conducting ability of graphene.\cite{yang12apl,li10prb} However, recent studies on corrugated CNTs\cite{ZhuHQ-12NJP} and CNT-GNR interfaces\cite{ChenXB-13PRB} find that a strong structural deformation, if the $sp^2$ bonding configuration is preserved, could be viewed as a perturbation to the transport of most low-frequency phonons, leading to a slight decrease in thermal conductance. The same conclusion should be also valid for graphene, which belongs to the $sp^2$ bonding systems as well.

In short summary, we have discussed various types of scattering events in graphene, including electron-phonon scattering, phonon-phonon scattering, defect scattering, and structural-distortion induced scattering. They all could be rather weak in graphene, due to the unique 2D planar structure and strong $sp^2$ bonding configuration. This explains the very long phonon MFP of graphene.

\section{Extrinsic Thermal Conductivity of Graphene}

The long phonon MFP in pristine graphene would suggest that it is possible to tune thermal conductivity more effectively by introducing extrinsic scattering mechanisms which dominate over intrinsic scattering mechanisms in graphene. For example, isotope scattering, normally unimportant with respect to other scattering processes, could become significant in graphene thermal conduction. It could be easier to observe size effects on thermal transport in graphene because samples do not have to be extremely shrunk. In the following, we discuss various scattering mechanisms and their influences on thermal conduction separately, giving rise to tunable extrinsic thermal conductivity of graphene.

\subsection{Isotope Effects}

\begin{figure} [tbp]
\centering
\includegraphics[width=0.4\textwidth]{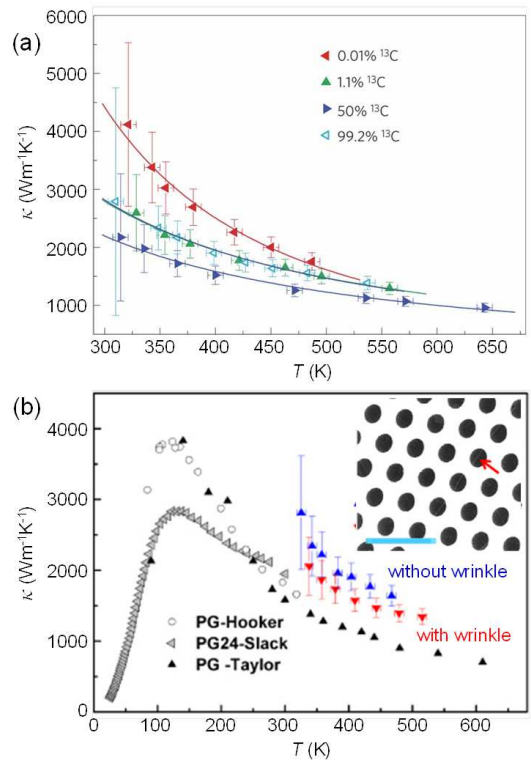}
\caption{\label{fig:kg-imperfect} a) Thermal conductivity of suspended CVD graphene as a function of temperature for different $^{13}$C concentrations, showing isotope effect. Reproduced with permission.\cite{ChenSS-12NM} Copyright 2012, NPG. b) Thermal conductivity of suspended CVD graphene with (red down-triangle) and without (blue up-triangle) wrinkles as a function of temperature. Also shown in comparison are the literature thermal conductivity data of pyrolytic graphite samples.\cite{Slack-62PR,Hooker-65PRSLA,Taylor-66PM} Inset shows the SEM image of CVD graphene on the Au-coated SiN$_{\rm x}$ holey membrane. The red arrow indicates a wrinkle. Scale bar is 10 $\mu$m. Reproduced with permission.\cite{ChenSS-12NaT} Copyright 2012, IOP.}
\end{figure}

The knowledge of isotope effects on thermal transport properties is valuable for tuning heat conduction in graphene. Natural abundance carbon materials are made up of two stable isotopes of $^{12}$C (98.9\%) and $^{13}$C (1.1\%). Changing isotope composition can modify dynamic properties of crystal lattices and affect their thermal conductivity.\cite{HuJN-10APL,Lindsay-13PRB} For instance, it has been found that at room temperature isotopically purified diamond has a thermal conductivity of $\sim$3300 Wm$^{-1}$K$^{-1}$,\cite{Anthony-90PRB,Berman-92PRB} about 50\% higher than that of natural diamond, $\sim$2200 Wm$^{-1}$K$^{-1}$.\cite{Ho-72JPCRD} Similar effects have also been observed in 1D nanostructures, boron nitride nanotubes.\cite{ChangCW-06PRL} Very recently, the first experimental work to show the isotope effect on graphene thermal conduction was reported by Chen et al.\cite{ChenSS-12NM} By using the CVD technique, they synthesized isotopically modified graphene containing various percentages of $^{13}$C. The graphene flakes were subsequently suspended over 2.8-$\mu$m-diameter holes and thermal conductivity was measured by the Raman thermometry technique. As shown in \textbf{Figure \ref{fig:kg-imperfect}a}, compared with natural abundance graphene (1.1\% $^{13}$C), the $\kappa$ values were enhanced in isotopically purified samples (0.01\% $^{13}$C), and reduced in isotopically mixed ones (50\% $^{13}$C).

Isotope effects influence thermal conduction in two aspects: (i) modify phonon dispersion and (ii) introduce isotope scattering. We will analyze the two contributions separately for graphene.

\begin{figure}
\includegraphics[width=\linewidth]{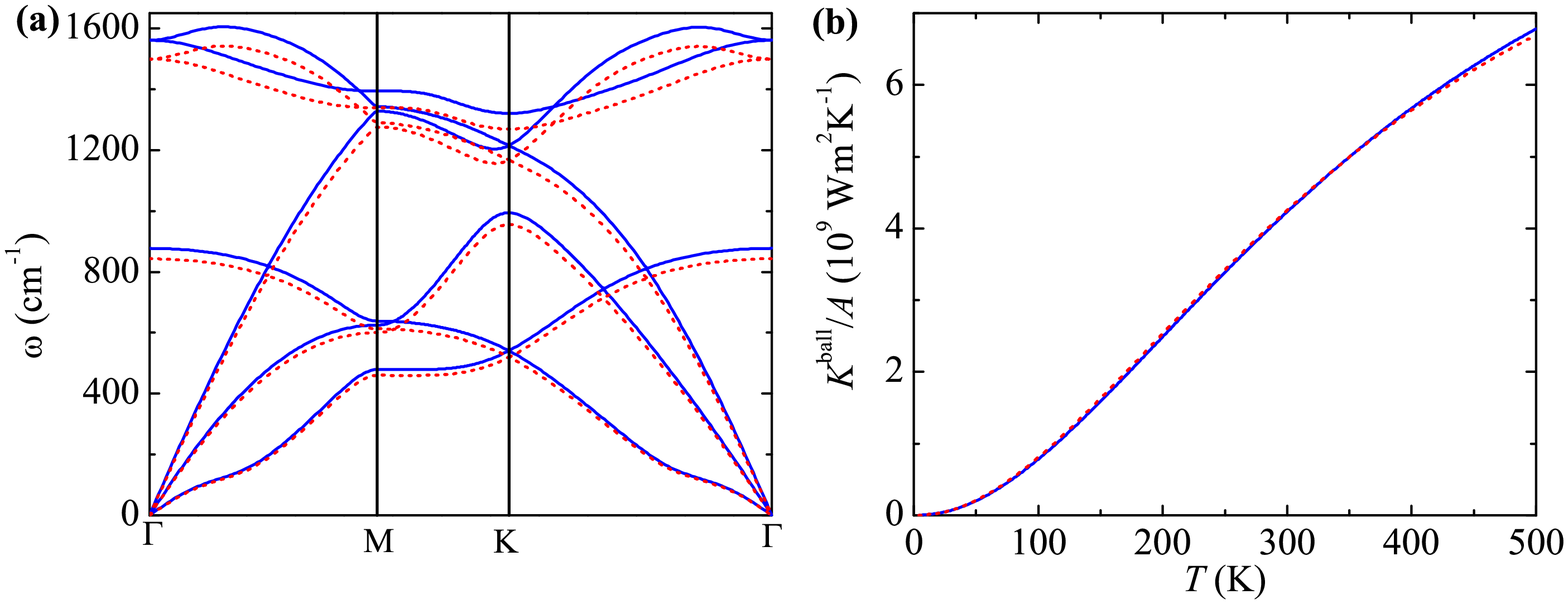}
\caption{\label{phonon-iso} a) Phonon dispersion and b) $K^{\rm{ball}} / A$ as a function of $T$ for graphene systems comprising isotopically pure $^{12}$C (blue solid line) and  $^{13}$C (red dashed line).}
\end{figure}

\textbf{Figure \ref{phonon-iso}a} shows phonon dispersions for graphene systems comprising isotopically pure $^{12}$C and $^{13}$C.  The replacement of $^{12}$C by $^{13}$C lowers frequencies of phonons. Such effects are negligible for low-frequency acoustic modes and become noticeable for high-frequency optical modes. In contrast, no observable frequency shift happens when varying the $^{13}$C isotope concentration from 0 to the natural abundance of 1.1\% (see Figure \ref{phonon-iso}a and Figure \ref{phonon}a).

The isotope-induced frequency shift can be detected by Raman experiments. Two types of Raman peaks are measured in graphene: the G peak corresponding to the doubly degenerated LO and TO mode at the $\Gamma$ point ($E_{\rm 2g}$ symmetry), and the 2D peak associated with the TO phonons nearby the K point.\cite{ferrari06prl,malard09pr} As shown in Figure \ref{phonon-iso}a, when changing the atomic mass $M$ from 12 to 13, the TO and LO modes at the $\Gamma$ point shift from 1561 to 1500 cm$^{-1}$, and the TO mode at the K point shifts from 1321 to 1269 cm$^{-1}$. It is known that the frequency of optical phonons at the $\Gamma$ point is proportional to $M^{-1/2}$. This tells us that the frequency would decrease by $1 - \sqrt{12/13} \approx 4\%$ when varying  $M$ from 12 to 13. Both calculations predict a red shift of 61 cm$^{-1}$ in the G peak, in good agreement with experiment (about 64 cm$^{-1}$).\cite{ChenSS-12NM}

The isotope-induced change in phonon dispersion may affect ballistic thermal conductance. To see the effects, we calculate $K^{\rm{ball}} / A$ as a function of $T$ for graphene with $M = 12$ and 13, and present the results in \textbf{Figure \ref{phonon-iso}b}. Interestingly, the two different $M$ give essentially the same $K^{\rm{ball}} / A$. This is easy to understand in the Landauer picture. The isotope-induced frequency shifts are small in magnitude and are limited to the high-frequency region. Both features, independent of materials types, result in minor changes in $K^{\rm{ball}} / A$. The unimportant isotope effect on $K^{\rm{ball}} / A$ should apply generally to other materials. This implies that the isotope-induced change in thermal conductivity comes from isotope scattering.

Isotope impurities as point defects are characterized by foreign atoms with $M_i$ different from host atoms. The variance in atomic mass causes phonon scattering, which is usually described as Rayleigh scattering. The scattering rate is $1/\tau_i \propto (\delta M)^2/ \lambda_{\rm w}^{\alpha}$, where $\delta M = 1 - M_i/\overline{M}$, $\overline{M}$ is the average atomic mass, $\lambda_{\rm w}$ is the phonon wavelength, and the exponent $\alpha = 3 (4)$ for 2D (3D) systems.\cite{klemens55,klemens94carbon,klemens00} The formula states that isotope scattering mainly influences short-wavelength phonons, keeping long-wavelength phonons unaffected. The isotope scattering has minor influence on thermal conduction at very low temperatures when only low-frequency acoustic phonons are thermally excited, as well as at very high temperatures when electron-phonon and phonon-phonon interactions dominate phonon scattering. The isotope scattering could affect thermal conduction significantly provided that other scattering processes are relatively weak. For graphene, we expect important isotope effects on thermal conductivity in high-quality samples of large grains at intermediate temperatures.

The above theoretical picture is well supported by the experiment by Chen et al.,~\cite{ChenSS-12NM} whose main results are shown in Figure \ref{fig:kg-imperfect}a. They measured graphene samples with the $^{13}$C isotope concentration $\rho$ = 0.01\%, 1.1\%, 50\% and 99.2\%, and found that the room-temperature $\kappa$ is about 4419, 2792, 2197 and 2816 Wm$^{-1}$K$^{-1}$, respectively. $\kappa$ of the natural abundance $\rho = 1.1\%$ is close to that of $\rho = 99.2\%$. While $\kappa$ is insensitive to the average atomic mass, it can be tuned largely by varying the concentration of isotope impurity. Compared to the natural abundance system, room-temperature $\kappa$ is lowered by 21\% when increasing $\rho$ to 50\%, and enhanced by 58\% when decreasing $\rho$ to 0.01\%. The enhancement becomes smaller with increasing temperature, as other scattering processes (like many-body scattering) become dominated. At $T \approx 450$ K, $\kappa$ of $\rho = 0.01\%$ is about 2000 Wm$^{-1}$K$^{-1}$, 25\% higher than that of $\rho = 1.1\%$.

Then we compare graphene with other materials. At room temperature, the isotopic enrichment induced enhancement of $\kappa$ in graphene (58\%) is comparable to those obtained in diamond (50\%)~\cite{Anthony-90PRB,Berman-92PRB} and boron nitride nanotubes (50\%),~\cite{ChangCW-06PRL} and larger than those achieved in silicon (10\%)~\cite{kremer04ssc} and germanium (30\%).~\cite{asen97prb} The large enhancement of diamond and boron nitride nanotubes, as for graphene, is presumably because the experimental samples have low defect concentrations. It is known that isotope scattering would be overridden by other defect scattering processes in samples with high defect concentrations. The observation of large isotope effects on thermal conductivity could serve as an indication of high sample quality.

\subsection{Structural Defect Effects}

Structural defects are common in fabricated graphene, especially in CVD grown graphene.\cite{Wood-11NaL,Koepke-13ACSN} The effects of wrinkles\cite{ChenSS-12NaT} and grain size\cite{Vlassiouk-11NaT} on the thermal conduction of suspended single-layer CVD graphene have been examined in experiments by using the Raman thermometry technique. Chen et al.\cite{ChenSS-12NaT} found that the thermal conductivity of graphene with obvious wrinkles (indicated by arrows in the inset of \textbf{Figure \ref{fig:kg-imperfect}b}) is about 15-30\% lower than that of wrinkle-free graphene over their measured temperature range, $\sim$330-520 K (Figure \ref{fig:kg-imperfect}b). Vlassiouk et al.\cite{Vlassiouk-11NaT} measured suspended graphene with different grain sizes obtained by changing the temperature of CVD growth. The grain sizes $\ell_{\rm G}$ were estimated to be 150 nm, 38 nm, and 1.3 nm in different samples in terms of the intensity ratio of the G peak to D peak in Raman spectra.\cite{Cancado-06APL} Since grain boundaries in graphene serve as extended defects and scatter phonons, graphene with smaller grain sizes are expected to suffer more frequent phonon scattering. Their measured thermal conductivity does show a decrease for smaller grain sizes, indicating the observation of the grain boundary effect on thermal conduction. Whereas, the dependence on the grain size shows a weak power law, $\kappa \sim \ell_{\rm G}^{1/3}$, for which there is no theoretical explanation yet.\cite{Vlassiouk-11NaT} However, for SiO$_2$-supported graphene, recent theoretical work based on NEGF method showed a similar but stronger dependence of $\kappa$ on the grain size $\ell_{\rm G}$ in the range of $\ell_{\rm G}<1~\mu$m.\cite{Serov-13APL} Further experimental studies are required to reveal the grain size effects on the thermal transport of both suspended and supported graphene.

Point defect scattering is usually described as Rayleigh scattering, which as shown above mainly affects transport of short-wavelength phonon modes. The influence of point defects on thermal transport has been extensively studied in theoretical works for various types of defects in graphene, including Stone-Wales defects,~\cite{hao11apl,haskins11an,xie2011thermal} substitutional defects,~\cite{mortazavi12ssc,jiang11apl} and vacancies.~\cite{zhang11prb,hao11apl,haskins11an,jiang11apl} In contrast, extended defects attracted less attention,~\cite{bagri11nl,lu12apl,Serov-13APL,HuangHQ-13PRB} although in experiments they ubiquitously exist probably caused by substrate imperfections or kinetic factors~\cite{coraux09njp,park10carbon} and can be precisely controlled nowadays.~\cite{lahiri10nn} Distinct from point defects, extended defects could induce a significant suppression of phonon transmission in a wide frequency range, leading to a large decrease in thermal conductance.~\cite{Serov-13APL,HuangHQ-13PRB} While the defect-induced thermal conductance reduction moderately changes with varying defect types, it sensitively depends on the orientation between the extended defect and the transport direction.~\cite{HuangHQ-13PRB}

\subsection{Substrate Effects in Supported Graphene}

For practical applications, graphene is usually attached to a substrate in electronic and optoelectronic devices, so it is important to understand substrate effects on thermal properties of supported graphene~\cite{OngZY-11PRB,chen12nanoscale,xu2012heat,guo2012substrate}. Seol et al.\cite{Seol-10S,Seol-10JHT} measured exfoliated SLG on a 300 nm thick SiO$_2$ membrane by using the micro-resistance thermometry technique with the suspended bridge platform (Figure \ref{fig:thermometry}c). The observed thermal conductivity is $\kappa\sim$600 Wm$^{-1}$K$^{-1}$ near room temperature (solid black circles in Figure \ref{fig:kg-T}a). This value is much lower than those reported for suspended SLG via the Raman thermometry technique, but is still relatively high compared with those of bulk silicon ($\sim$150 Wm$^{-1}$K$^{-1}$) and copper ($\sim$400 Wm$^{-1}$K$^{-1}$). Another study by Cai et al.\cite{CaiWW-10NaL} showed CVD SLG supported on Au also has a decreased thermal conductivity, $\sim$370 Wm$^{-1}$K$^{-1}$ (this lower value compared to $\sim$600 Wm$^{-1}$K$^{-1}$ could be caused by grain boundary scattering in CVD graphene\cite{Serov-13APL}). The thermal conductivity reduction in supported graphene is attributed to substrate scattering, which strongly affects the out-of-plane flexural (ZA) mode of graphene.\cite{Seol-10S,OngZY-11PRB,QiuB-12APL} This effect becomes stronger in encased graphene, where graphene is sandwiched between bottom and top SiO$_2$. The thermal conductivity of SiO$_2$-encased exfoliated SLG was measured to be below 160 Wm$^{-1}$K$^{-1}$, reported by Jang et al.\cite{Jang-10NaL} using the micro-resistance thermometry technique with the substrate-supported platform (Figure \ref{fig:thermometry}e). For encased graphene, besides the phonon scattering by bottom and top oxides, the evaporation of top oxide could cause defects in graphene, which can further lower thermal conductivity. Knowing encased graphene $\kappa$ is useful for analyzing heat dissipation in top-gated graphene devices. Similar to suspended graphene, the intrinsic phonon MFP for supported graphene can be estimated based on Equation (\ref{kappa_l}). Since the graphene flakes measured by Seol et al.\cite{Seol-10S,Seol-10JHT} are very long ($\sim$10 $\mu$m) and relatively wide ($1.5-3.2~\mu$m), approaching the diffusive regime, their values could be treated as $\kappa^{\rm diff}$ for supported graphene. Combined with $K^{\rm ball}/A\approx4.2$ GWm$^{-2}$K$^{-1}$ at 300 K,\cite{XuY-09APL,Bae-13NC,Serov-13APL} the backscattering $\lambda$ of supported graphene is about 140 nm (or $\sim$ 90 nm for common MFP).\cite{Bae-13NC}

\subsection{Size Effects and Boundary Scattering}

In macroscopic bulk materials, thermal conductance satisfies Fourier's scaling law in the diffusive region, $K = \kappa A/L$, where $\kappa$ is an intrinsic material property, independent of system size. The scaling law could break down in nanostructures caused by two mechanisms: (i) thermal transport is not diffusive; and (ii) boundary effects become important.

For non-diffusive transport, $\kappa$ becomes length dependent. It is proportional to $L$ as $\kappa^{\rm ball}=(K^{\rm ball}/A)L$ in the ballistic limit, gets saturated in the diffusive limit, and typically grows gradually with increasing $L$ in the intermediate region. When quantum interference effects become important, phonons may transport in the localization region, accompanied by an exponentially decayed transmission. The localization effects, however, are difficult to observe in experiments, since thermal conduction is contributed by many phonon modes, whose localization lengths are strongly frequency dependent.~\cite{Savic-08PRL}

\begin{figure}
\includegraphics[width=0.75\linewidth]{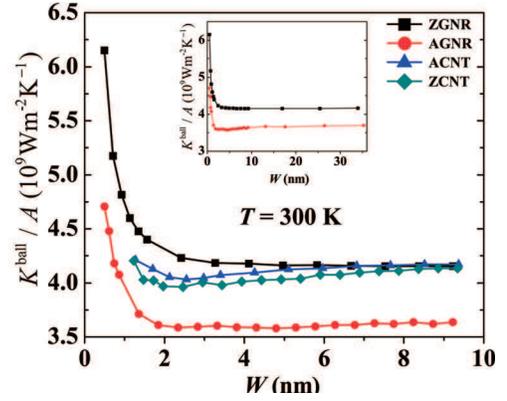}
\caption{\label{GNRs-APL} The ballistic thermal conductance per unit area $K^{\rm{ball}}/A$ at 300 K versus width ($W$) for zigzag GNRs (ZGNRs), armchair GNRs (AGNRs), armchair CNTs (ACNTs) and zigzag CNTs (ZCNTs). The inset shows $K^{\rm{ball}}/A$ for ZGNRs and AGNRs with
the width varying from 0.5 to 35 nm.  The lines are drawn to guide the eyes. Reproduced with permission.\cite{XuY-09APL} Copyright 2009, AIP. }
\end{figure}

In nanostructures, boundary effects cannot be neglected, and $\kappa$ gets dependent on $A$. $K^{\rm{ball}} / A$ is no longer constant as in the bulk but dependent on $A$ due to boundary effects, and $\lambda$ also depends on $A$ because the strength of boundary scattering varies with $A$. A previous work systematically studied $K^{\rm{ball}} / A$ of GNRs with different ribbon widths $W$ and edge shapes using the NEGF method.~\cite{XuY-09APL} The main results are shown in \textbf{Figure \ref{GNRs-APL}}. As $W$ increases, $K^{\rm{ball}} / A$ has a signifiant decrease when $W < 2$ nm and changes slowly with $W$ when $W > 2$ nm. The room-temperature $K^{\rm{ball}} / A$ of GNRs with zigzag edges is about $4.2 \times 10^9$ Wm$^{-2}$K$^{-1}$, close to those of graphene and CNTs.\cite{Mingo-05PRL} In contrast, GNRs with armchair edges have obviously lower $K^{\rm{ball}} / A$, displaying an interesting anisotropy in ballistic thermal conduction. The anisotropy can be as large as 30\%, which could be further enhanced by modifying the edge shape.~\cite{HuangHQ-13PRB} The anisotropic $K^{\rm{ball}} / A$ generally exists in narrow GNRs and is expected to disappear when $W > 100$ nm. Considering that graphene, as their bulk counterpart, is isotropic in thermal conduction (basal-plane), such an anisotropy obviously comes from boundary effects, which could be very long-range for phonon transport. We expect similar anisotropy in other materials. Anisotropic ballistic thermal conduction is also found in silicon nanowires (SiNWs) but was explained by the anisotropy of the bulk phonon dispersion.~\cite{markussen08nl}

In practice edges of graphene samples are not atomically regular but rather rough. The edge roughness causes phonon scattering and decreases $\kappa$. The phonon MFP associated with boundary scattering generally is described as $\lambda_{\rm B} = D (1+p)/(1-p)$.~\cite{ziman2001,nika09prb,Balandin-11NM} $D$ is the dimension of the sample perpendicular to the transport direction, and $D = W$ for GNRs here. $p$ is an empirical parameter defined as the probability of specular scattering at the boundary. In the extremely rough limit ($p = 0$), $\lambda_{\rm B} = D$, corresponding to fully diffusive scattering at the boundary. In general, $p$ is determined by the roughness of boundary and is between 0 and 1.

Boundary scattering could be a dominated scattering process in nanostructures, although it is usually unimportant in large samples. Silicon serves as an excellent example on this aspect. Bulk silicon has a large room-temperature $\kappa$ of 150 Wm$^{-1}$K$^{-1}$ and $\lambda \sim 150$ nm as shown above. When nanostructuring silicon into nanowires, $\kappa$ decreases seriously with respect to the bulk counterpart by about 100 fold.~\cite{LiDY-03APL,Hochbaum-08N,Boukai-08N} As $K^{\rm{ball}} / A$ is very similar between bulk~\cite{jeong11jap} and nanowires,~\cite{markussen08nl} SiNWs actually have $\lambda$ about 100-fold smaller than the bulk. The results are quite amazing. By simply nanostructuring the material, thermal conductivity can be lowered orders of magnitude. This has strong implications to thermoelectrics. For instance, the thermoelectric figure of merit improves by 100 times through nanostructuring silicon into SiNWs, mainly contributed by the decrease of $\kappa$.~\cite{Hochbaum-08N,Boukai-08N}

\begin{figure} [tbp]
\centering
\includegraphics[width=0.49\textwidth]{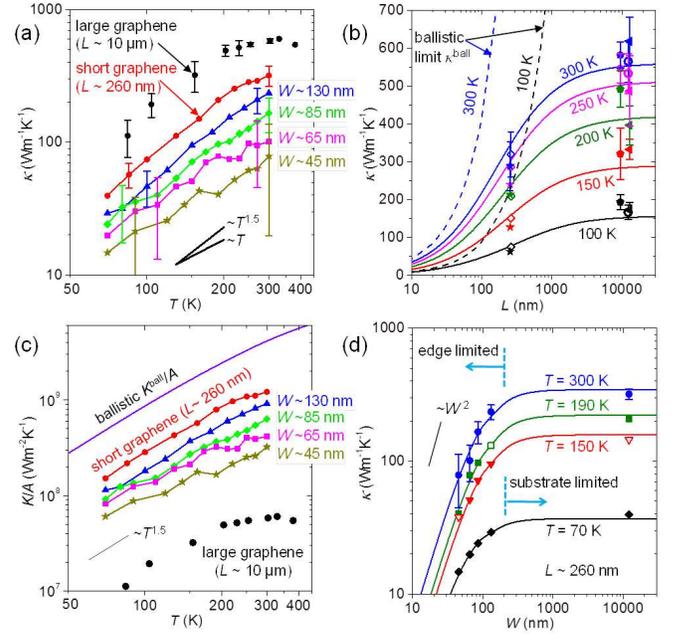}
\caption{\label{fig:kg-L-W} a) Thermal conductivity vs. temperature for SiO$_2$-supported GNRs\cite{Bae-13NC} ($L\approx260$ nm, $W$ as listed), ``short" but wide graphene\cite{Bae-13NC} ($L\approx260$ nm, $W\approx12~\mu$m), and ``large" graphene\cite{Seol-10S} ($L\approx10~\mu$m, $W\approx2.4~\mu$m). b) Thermal conductivity reduction with length for ``wide" graphene ($W\gg\lambda$), compared to the ballistic limit ($\kappa^{\rm ball} = K^{\rm ball}L/A$) at several temperatures. Symbols are data from ``short" and ``large" samples. Solid lines are model from Equation (\ref{eq:kL}). c) Thermal conductance per unit cross-sectional area ($K/A$) for the same samples as in a). The short but wide graphene sample attains up to ~35\% of the theoretical ballistic heat flow limit, $K^{\rm ball}/A$. d) Thermal conductivity reduction with width for GNRs, all with $L\approx260$ nm. Solid symbols are experimental data from a), open symbols are interpolations for the listed temperature; lines are fitted model from Equation (\ref{eq:kW}). Reproduced with permission.\cite{Bae-13NC} Copyright 2013, NPG.}
\end{figure}

By developing the substrate-supported thermometry platform (Figure \ref{fig:thermometry}f), Bae et al.\cite{Bae-13NC} for the first time measured the size effects on thermal transport in SiO$_2$-supported, exfoliated graphene whose length and width are comparable to phonon MFP ($\lambda\sim$ 140 nm as shown above). The obtained room temperature $\kappa$ of 260-nm-long graphene is $\sim$320 Wm$^{-1}$K$^{-1}$, obviously lower than that of $\sim$10-$\mu$m-long graphene reported by Seol et al.\cite{Seol-10S} (\textbf{Figure \ref{fig:kg-L-W}a}). The data from these two studies can be captured by the simple model of length-dependent thermal conductivity\cite{Bae-13NC,Munoz-10NaL}
\begin{equation}\label{eq:kL}
    \kappa(L)=[\frac{1}{(K^{\rm ball}/A)L}+\frac{1}{\kappa^{\rm diff}}]^{-1},%=\frac{K^{\rm ball}}{A}(\frac{1}{L}+\frac{1}{\lambda_{\rm bs}})^{-1}
\end{equation}
with choices of simulated $\kappa^{\rm diff}$ for SiO$_2$-supported graphene by Seol et al.\cite{Seol-10S} (see \textbf{Figure \ref{fig:kg-L-W}b}). As shown in \textbf{Figure \ref{fig:kg-L-W}c}, the corresponding thermal conductance per unit cross-sectional area, $K/A=\kappa/L$ reaches $\sim$30-35\% of theoretical ballistic upper limit $K^{\rm ball}/A$ up to room temperature for 260-nm-long graphene, indicating the quasi-ballistic transport ($L\sim\lambda$). The ballistic percentage is consistent with a simple estimation of transmission probability, $\lambda/(\lambda+L)$, using sample length $L$ and back-scattering MFP $\lambda\sim140$ nm for SiO$_2$-supported graphene. This is the first time of observing quasi-ballistic thermal transport at \emph{room temperature} in any materials, enabled by the relatively large phonon MFP of graphene. Previous observations of ballistic thermal transport were only at very low temperature ($\sim$1 K).\cite{Schwab-00N}

Bae et al.\cite{Bae-13NC} further etched short ($L\sim$260 nm) graphene into arrays of nanoribbons (Figure \ref{fig:thermometry}f), whose width $W\approx$ 45-130 nm, comparable to $\lambda$ ($\sim$140 nm). The observed thermal conductance changes back to the diffusive regime gradually as the GNR width decreases from $\sim$130 nm to $\sim$45 nm (Figure \ref{fig:kg-L-W}c), correspondingly $\kappa$ decreases from 320 to 80 Wm$^{-1}$K$^{-1}$ at room temperature (Figure \ref{fig:kg-L-W}a), due to increased edge scattering in narrower GNRs. Experimental $\kappa$ versus $W$ can be fitted with an empirical model\cite{Bae-13NC}
\begin{equation}\label{eq:kW}
    \kappa_{\rm eff}(W,L)=[\frac{1}{c}(\frac{\Delta}{W})^n+\frac{1}{\kappa(L)}]^{-1},
\end{equation}
where $\Delta$ is the root-mean-square (r.m.s.) edge roughness and $\kappa(L)$ is given by Equation (\ref{eq:kL}) and Figure \ref{fig:kg-L-W}b. As shown by the solid lines in \textbf{Figure \ref{fig:kg-L-W}d}, the best-fit exponent $n$ for all listed temperatures is $1.8\pm0.3$, and this nearly $W^2$ dependence in edge-limited regime is consistent with previous findings for rough nanowires.\cite{ChenRK-08PRL,Martin-09PRL,Martin-10NaL,Sadhu-11PRB} The simple model appears to be a good approximation in a regime with $\Delta \ll W$, but it is likely to change in a situation with extremely rough edges, where the roughness correlation length\cite{Lim-12NaL} and phonon localization\cite{WangY-12APL} could also play an important role. Thus, deeper understandings of $W$ and $\Delta$ scalings due to edge-roughness scattering still require further theoretical studies and experimental observations. It is worth noting that through the electrical breakdown measurement on GNRs derived from unzipped CNTs,\cite{Kosynkin-09N,JiaoLY-09N,HuangB-09JACS,HuangB-10JCP} Liao et al.\cite{LiaoA-11PRL} was able to estimate GNR $\kappa$, and their values are slightly higher than those of Bae et al.\cite{Bae-13NC} for similar widths. Considering that CNT-unzipped GNRs have smoother edges,\cite{Kosynkin-09N,JiaoLY-09N} the two studies are essentially consistent.

\subsection{Interlayer Effects in Few-Layer Graphene}
\label{few-layer}

Interlayer scattering as well as top and bottom boundary scattering could take place in few-layer graphene, which could be another mechanism to modulate graphene thermal conductivity. It is interesting to investigate the evolution of the thermal conductivity of FLG with increasing thickness, denoted by the number of atomic layers ($n$), and the critical thickness needed to recover the thermal conductivity of graphite.

\begin{figure} [tbp]
\centering
\includegraphics[width=0.4\textwidth]{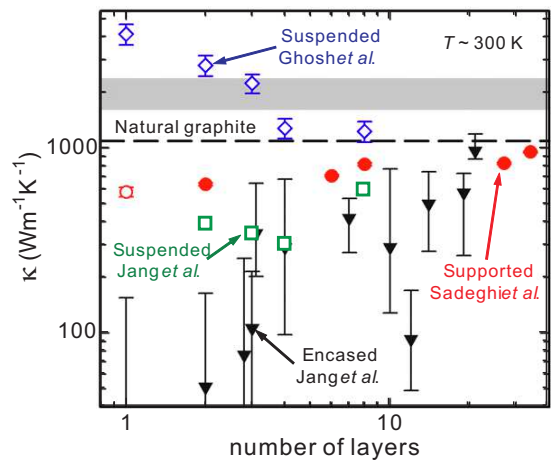}
\caption{\label{fig:kg-N} Experimental in-plane thermal conductivity near room temperature as a function of the number of layers $n$ for suspended graphene by Ghosh et al.\cite{Ghosh-10NM} (open blue diamond) and by Jang et al.\cite{Jang-13APL} (open green square), SiO$_2$-supported graphene by Seol et al.\cite{Seol-10S} (open red circle) and Sadeghi et al.\cite{Sadeghi-13PNAS} (solid red circle), and SiO$_2$-encased graphene by Jang et al.\cite{Jang-10NaL} (solid black triangle). The data show a trend to recover the value (dashed line) measured by Sadeghi et al.\cite{Sadeghi-13PNAS} for natural graphite source used to exfoliate graphene. The gray shaded area shows the highest reported $\kappa$ values of pyrolytic graphite.\cite{Slack-62PR,Hooker-65PRSLA,Taylor-66PM} Reproduced with permission.\cite{Sadeghi-13PNAS} Copyright 2013, NAS.}
\end{figure}

Several experimental studies on this topic have been conducted for encased,\cite{Jang-10NaL} supported,\cite{Sadeghi-13PNAS} and suspended FLG,\cite{Ghosh-10NM,Jang-13APL} and their results are summarized in \textbf{Figure \ref{fig:kg-N}}. Jang et al.\cite{Jang-10NaL} measured the thermal transport of SiO$_2$-encased FLG by using the substrate-supported, micro-resistance thermometry platform (Figure \ref{fig:thermometry}e). They found that the room-temperature thermal conductivity increases from $\sim$50 to $\sim$1000 Wm$^{-1}$K$^{-1}$ as the FLG thickness increases from 2 to 21 layers, showing a trend to recover natural graphite $\kappa$. This strong thickness dependence was explained by the top and bottom boundary scattering and disorder penetration into FLG induced by the evaporated top oxide.\cite{Jang-10NaL} Very recently, another similar yet less pronounced trend was observed in SiO$_2$-supported FLG by Sadeghi et al.\cite{Sadeghi-13PNAS} using the suspended micro-resistance thermometry platform (similar to Figure \ref{fig:thermometry}c). As shown by red dots in Figure \ref{fig:kg-N}, the measured room-temperature $\kappa$ increases slowly as increasing thickness, and the recovery to natural graphite would occur even more than 34 layers. The difference between the results by Jang et al.\cite{Jang-10NaL} and Sadeghi et al.\cite{Sadeghi-13PNAS} is not unexpected, because encased FLG $\kappa$ could be suppressed much more in thin layers than thick layers due to the effect of top oxide, and hence shows a stronger thickness dependence.

For suspended FLG, there are two contradictory observations in the thickness dependence. At first, based on the Raman thermometry technique (Figure \ref{fig:thermometry}a) Ghosh et al.\cite{Ghosh-10NM} showed a decrease of suspended FLG $\kappa$ from the SLG high value to regular graphite value as thickness increases from 2 to 8 layers (open diamonds in Figure \ref{fig:kg-N}). The $\kappa$ reduction was explained by the interlayer coupling and increased phase-space states available for the phonon Umklapp scattering in thicker FLG.\cite{Ghosh-10NM} However, a very recent study by Jang et al.\cite{Jang-13APL} seems to show a different thickness trend for suspended FLG. They measured thermal conductivity of suspended graphene of 2-4 and 8 layers by using a modified T-bridge micro-resistance thermometry technique. The obtained room-temperature $\kappa$ for 2-4 layers is about 300-400 Wm$^{-1}$K$^{-1}$ with no apparent thickness dependence, while $\kappa$ for 8-layer shows an increase to $\sim$600 Wm$^{-1}$K$^{-1}$ (open squares in Figure \ref{fig:kg-N}). Surprisingly, this trend is qualitatively in agreement with that of Sadeghi et al.\cite{Sadeghi-13PNAS} for \emph{supported} FLG; both show similar increasing amounts of $\kappa$ from 2 to 8 layers (Figure \ref{fig:kg-N}), despite a small decrease from 2 to 4 layers in the former, which could arise from different sample qualities and measurement uncertainty. Given opposite thickness trends of Ghosh et al.\cite{Ghosh-10NM} and Jang et al.,\cite{Jang-13APL} further experimental works are required to clarify the real thickness-dependent $\kappa$ in \emph{suspended} FLG. Moreover, we want to point out that suspended FLG $\kappa$ values of Jang et al.\cite{Jang-13APL} are close to those reported by Pettes et al.\cite{Pettes-11NaL} for suspended bilayer graphene (BLG), $\sim$600 Wm$^{-1}$K$^{-1}$ at room temperature (see Figure \ref{fig:kg-T}a). Both are much lower than suspended SLG $\kappa$. The latter attributed this to phonon scattering by a residual polymeric layer on graphene,\cite{Pettes-11NaL} even though the former claimed that electrical current annealing was used to remove polymer residues.\cite{Jang-13APL}

\subsection{Cross-Plane Thermal Conduction}

A remarkable feature of graphite and graphene is that their thermal properties are highly anisotropic. Despite high thermal conductivity along the in-plane direction, heat flow along the cross-plane direction (\emph{c} axis) is hundreds of times weaker, limited by weak van der Waals interactions between layers (for graphite) or with adjacent materials (for graphene). For example, the thermal conductivity along the \emph{c} axis of pyrolytic graphite is only $\sim$6 Wm$^{-1}$K$^{-1}$ at room temperature\cite{Ho-72JPCRD} (Figure \ref{fig:kg-T}a). For graphene, it is often attached to a substrate or embedded in a medium for potential applications. Heat conduction along the cross-plane direction is characterized by the thermal interface/boundary conductance between graphene and adjacent materials, which could become a limiting dissipation bottleneck in highly scaled graphene devices and interconnects.\cite{Pop-12MRSB,Bae-10NaL,Bae-11ACSN,Behnam-12NaL,Islam-13EDL}

\begin{figure} [tbp]
\centering
\includegraphics[width=0.4\textwidth]{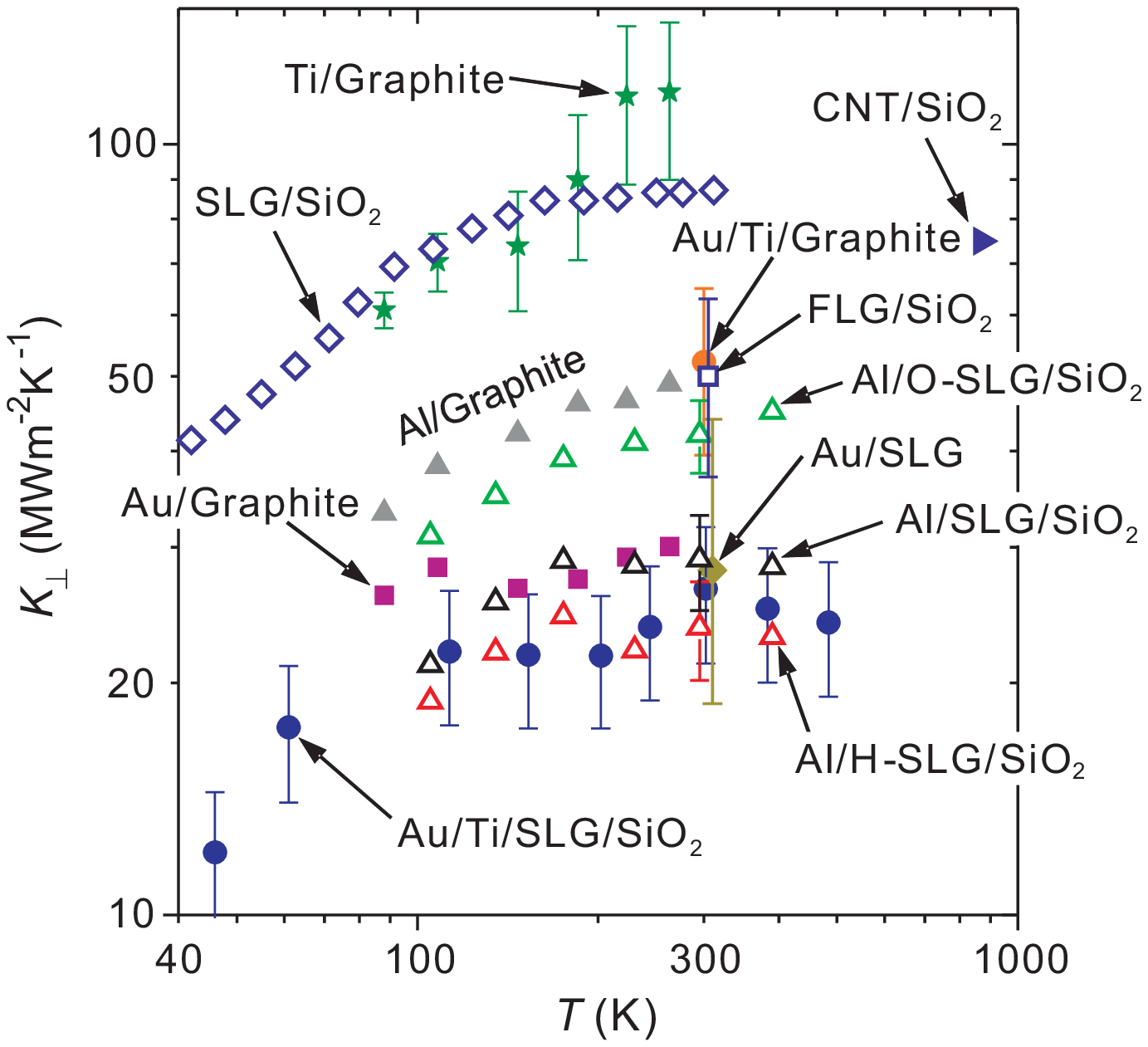}
\caption{\label{fig:G-interface} Experimental thermal interface conductance $K_{\perp}$ vs. temperature for SLG/SiO$_2$ by Chen et al.\cite{ChenZ-09APL} (open purple diamond), FLG/SiO$_2$ by Mak et al.\cite{Mak-10APL} (open purple square), CNT/SiO$_2$ by Pop et al.\cite{Pop-07JAP} (solid purple right-triangle), Au/SLG by Cai et al.\cite{CaiWW-10NaL} (solid gold diamond), Au/Ti/SLG/SiO$_2$ (solid blue circle) and Au/Ti/graphite (solid orange circle) by Koh et al.,\cite{Koh-10NaL} interfaces of graphite with Au (solid magenta square), Al (solid gray up-triangle), Ti (solid green asterisk) by Schmidt et al.,\cite{Schmidt-10JAP} interfaces of Al/SLG/SiO$_2$ without treatment (open black up-triangle), with oxygen treatment (Al/O-SLG/SiO$_2$, open green up-triangle), and with hydrogen treatment (Al/H-SLG/SiO$_2$, open red up-triangle) by Hopkins et al.\cite{Hopkins-12NaL} }
\end{figure}

The thermal interface conductance across graphene/graphite and other materials has been measured by using 3$\omega$ method,\cite{ChenZ-09APL} time-domain thermoreflectance (TDTR) technique,\cite{Mak-10APL,Schmidt-10JAP,Norris-12JHT,Koh-10NaL,Hopkins-12NaL,ZhangCW-13Carbon} and Raman-based method,\cite{CaiWW-10NaL,Ermakov-13NaT} Most experimental data available to date are shown in \textbf{Figure \ref{fig:G-interface}}, and they are consistent with each other in general, given the variations of sample qualities and measurement techniques. Chen et al.\cite{ChenZ-09APL} and Mak et al.\cite{Mak-10APL} showed the thermal interface conductance per unit area of graphene/SiO$_2$ is $K_{\perp}\sim50-100$ MWm$^{-2}$K$^{-1}$ at room temperature, with no strong dependence on the FLG thickness. Their values are close to that of CNT/SiO$_2$,\cite{Pop-07JAP} reflecting the similarity between graphene and CNT. Schmidt et al.\cite{Schmidt-10JAP} measured $K_{\perp}$ of the graphite/metal interfaces, including Au, Cr, Al, and Ti. Among them, the graphite/Ti has the highest $K_{\perp}$, $\sim$120 MWm$^{-2}$K$^{-1}$, and the graphite/Au interface has the lowest $K_{\perp}$, $\sim$30 MWm$^{-2}$K$^{-1}$ near room temperature. Their $K_{\perp}$ of graphite/Au is consistent with the value by Norris et al.\cite{Norris-12JHT} and values of SLG/Au by Cai et al.\cite{CaiWW-10NaL} and FLG/Au by Ermakov et al.\cite{Ermakov-13NaT} Koh et al.\cite{Koh-10NaL} later measured heat flow across the Au/Ti/$n$-LG/SiO$_2$ interfaces with the layer number $n=1-10$. Their observed room-temperature $K_{\perp}$ is $\sim$25 MWm$^{-2}$K$^{-1}$, which shows a very weak dependence on the layer number $n$ and is equivalent to the total thermal conductance of Au/Ti/graphite and graphene/SiO$_2$ interfaces acting in series. This indicates that the thermal resistance of two interfaces between graphene and its environment dominates over that between graphene layers. Interestingly, Hopkins et al.\cite{Hopkins-12NaL} showed the thermal conduction across the Al/SLG/SiO$_2$ interface could be manipulated by introducing chemical adsorbates between the Al and SLG. As shown in Figure \ref{fig:G-interface}, their measured $K_{\perp}$ of untreated Al/SLG/SiO$_2$ is $\sim$30 MWm$^{-2}$K$^{-1}$ at room temperature, in agreement with Zhang et al.\cite{ZhangCW-13Carbon} The $K_{\perp}$ increases to $\sim$42 MWm$^{-2}$K$^{-1}$ for oxygen-functionalized graphene (O-SLG), while decreases to $\sim$23 MWm$^{-2}$K$^{-1}$ for hydrogen-functionalized graphene (H-SLG). These effects were attributed to changes in chemical bonding between the metal and graphene, and are consistent with the observed enhancement in $K_{\perp}$ from the Al/diamond\cite{Stoner-92PRL} to Al/O-diamond interfaces.\cite{Collins-10APL}

\section{Thermoelectric Properties of Graphene}

Thermoelectric materials can convert waste heat into electricity by the Seebeck effect and use electricity to drive electronic cooling or heating by the Peltier effect. Thermoelectric devices are all-solid-state devices with no moving part, thus are silent, reliable and scalable. However, they only find limited applications due to their low efficiency. The efficiency of a thermoelectric material is determined by the thermoelectric figure of merit ($ZT$), which typically is defined as~\cite{goldsmid1986}
\begin{equation}
ZT=\sigma S^2 T/\kappa,
\end{equation}
where $\sigma$ is the electrical conductivity, $S$ is the Seebeck coefficient [also called thermoelectric power (TEP) or thermopower], $T$ is the absolute temperature, and the thermal conductivity $\kappa = \kappa_e + \kappa_l$ have contributions from electrons ($\kappa_e$) and lattice vibrations ($\kappa_l$). $\kappa_e$ is usually extracted based on the Wiedemann-Franz Law $\kappa_e / \sigma =  L_0 T$, where the Lorenz number $L_0$ is equal to $2.44 \times 10^{-8}$ W$\Omega$K$^{-2}$ for free electrons. This law does not always hold. For example, $\kappa_e$ becomes zero for a delta-shaped transport distribution.~\cite{mahan1996best} However, in graphene $\kappa_e$ is negligible with respect to $\kappa_l$,~\cite{Saito-07PRB,Watanabe-09PRB,Balandin-08NaL,Fong-13PRX,Yigen-13PRB,Yigen-14NaL} similar as in CNTs.~\cite{Hone-99PRB,Kim-01PRL} Currently, the state-of-art commercial thermoelectric materials, like Bi$_2$Ti$_3$, have room-temperature $ZT$ around 1.~\cite{Snyder-08NM}

Two important concepts have been developed to guide thermoelectrics research in the last twenty years.~\cite{Dresselhaus-07AM,Snyder-08NM} One was proposed by Hicks and Dresselhaus in 1993, which suggests to improve $ZT$ by going to lower dimensions.~\cite{hicks1993-2D,hicks1993-1D} Based on this concept, significant enhancements of $ZT$ to larger than 1 are found, for instance, in thin films~\cite{venkatasubramanian01science} and in quantum dot superlattices.~\cite{harman02science} The other concept, demonstrated by two seminal experiments in 2008,~\cite{Hochbaum-08N,Boukai-08N} suggests to increase $ZT$ by nanostructuring. The experiments found a 100-fold increase of $ZT$ in SiNWs over the bulk counterpart, mainly attributed to the decrease of $\kappa$ induced by boundary scattering. Graphene nanostructures can naturally combine the two concepts and might be useful for thermoelectrics research and applications.

For realizing large $ZT$ in graphene systems, two major disadvantages have to be overcome: (i) $\kappa$ is too high; (ii) $S$ is too small due to the gapless band structure. In the above section, we discussed various approaches to tune thermal conduction in graphene. Among them, boundary scattering, that can largely decrease $\kappa$ as demonstrated in SiNWs, is a promising way. Previous theoretical calculations predict that GNRs with disordered edge structures~\cite{sevinccli10prb} or graphene quantum dots~\cite{XuY-10PRB} may support large $ZT$. In addition, introducing weakly coupled interfaces can block thermal conduction efficiently. Taking CNTs as an example, despite extremely large $\kappa$ of single CNTs, when they are formed into random networks,\cite{Estrada-11APL,Behnam-12ACSN,Timmermans-12NaR,Gupta-12JAP,Gupta-13NaT} theoretical simulations predicted ultra small $\kappa$ of 0.13-0.2 Wm$^{-1}$K$^{-1}$ for networks due to junctions between CNTs.~\cite{prasher09prl} The same concept could be used for graphene to reduce $\kappa$. Next, we mainly discuss the Seebeck effect in graphene.

\begin{figure} [tbp]
\centering
\includegraphics[width=0.45\textwidth]{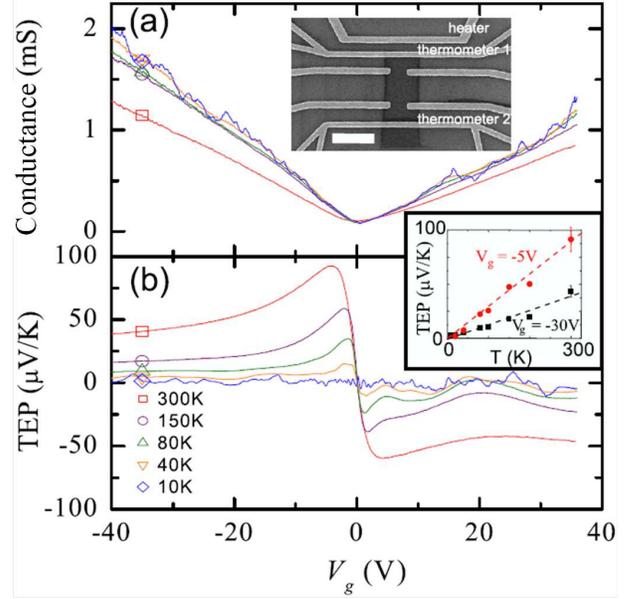}
\caption{\label{fig:thermopower} a) Electrical Conductance $G$ and b) thermopower TEP of a graphene sample as a function of back gate voltage $V_{\rm g}$ for $T$ = 300 K (square), 150 K (circle), 80 K (up triangle), 40 K (down triangle), and 10 K (diamond). Upper inset: SEM image of a typical device for thermoelectric measurements, scale bar is 2 $\mu$m. Lower inset: TEP values taken at $V_{\rm g}=-$30 V (square) and $-5$ V (circle). Dashed lines are linear fits to the data. Reproduced with permission.\cite{Zuev-09PRL} Copyright 2009, APS.}
\end{figure}

Thermoelectric transport in graphene has been experimentally investigated in the past five years.\cite{Zuev-09PRL,WeiP-09PRL,Checkelsky-09PRB,WangCR-10PRB,Nam-10PRB,Nam-13PRL,WangDQ-11PRB,LiuXF-12SSC,WuXS-11APL,Babichev-13JAP,XuXD-10NaL,Grosse-11NN} The Seebeck coefficient $S$ and electrical conductance $G$ of graphene can be measured against the gate voltage $V_{\rm g}$ (thus, carrier density $n_{\rm c}$) simultaneously via a widely-used microfabricated structure (inset of \textbf{Figure \ref{fig:thermopower}a}), which was developed by Small et al.\cite{Small-03PRL} to measure thermoelectric transport in CNTs. \textbf{Figure \ref{fig:thermopower}} shows typical results of measured $G$ and $S$ as a function of $V_{\rm g}$ in graphene at different temperatures.\cite{Zuev-09PRL} The Seebeck coefficient $S$ shows two peaks near the Dirac point (charge neutrality point) and changes its sign across the Dirac point as the majority carrier switches from electron to hole. The room-temperature peak values of $S$ for SLG and BLG are observed to be $\sim$50-100 $\mu$VK$^{-1}$ in different experiments.\cite{Zuev-09PRL,WeiP-09PRL,Checkelsky-09PRB,WangCR-10PRB,Nam-10PRB} For high carrier density $n_{\rm c}$ (i.e., high $|V_{\rm g}|$), the measured Seebeck coefficient scales as $S\sim1/\sqrt{|n_{\rm c}|}$ for SLG due to its linear dispersion,\cite{WeiP-09PRL} while $S\sim1/|n_{\rm c}|$ for BLG due to its hyperbolic dispersion,\cite{WangCR-10PRB} consistent with theories. Importantly, the simultaneous measurements of $G$ and $S$ enable testing the validation of the semiclassical Mott relation:\cite{Cutler-69PR}
\begin{equation}\label{Mott}
    S=-\frac{\pi^2k^2_{\rm B}T}{3|e|}\frac{1}{G}\frac{dG}{dV_{\rm g}}\frac{dV_{\rm g}}{dE}|_{E=E_{\rm F}},
\end{equation}
where $k_{\rm B}$ is the Boltzmann constant, $e$ is the electron charge, and $E_{\rm F}$ is the Fermi energy. For SLG the measured $S$ shows a linear $T$ dependence (inset of Figure \ref{fig:thermopower}b) and matches calculated $S$ from measured $G$ by Equation (\ref{Mott}),\cite{Zuev-09PRL,WeiP-09PRL,Checkelsky-09PRB} indicating an agreement with the Mott relation. For BLG, however, the agreement only holds for high carrier density; for low carrier density there is an obvious difference between measured and calculated $S$ as well as a deviation from the linear $T$ dependence at high temperature.\cite{WangCR-10PRB,Nam-10PRB} This failure of the Mott relation was attributed to the low Fermi temperature in BLG.

The thermoelectric properties of materials can be also probed by using a conducting tip to measure the thermoelectric voltage between the sample and tip, induced by a given temperature difference between them. By employing atomic force microscopy (AFM) and scanning tunneling microscopy (STM) techniques, Cho et al.\cite{Cho-13NM} and Park et al.\cite{ParkJ-13NaL} measured the thermopower of epitaxial graphene on SiC, respectively. The advantage of this method is the simultaneous imaging of the sample structure and thermoelectric signals with a spatial resolution of atomic-scale. Since the Seebeck coefficient relies on the sample local density of states (LDOS) near the Fermi energy, and LDOS can be quite different in the presence of boundaries and disorders\cite{OuYangFP-08JPCC,HuangB-08JPCC,LiJ-10PRB,LinF-09CPL,li2012dirac}, thermoelectric imaging allows us to probe grain boundaries, wrinkles, defects, and impurities in graphene, which may not be reflected in topography images.\cite{Cho-13NM,ParkJ-13NaL}

For practical applications, the Seebeck coefficient and power factor $\sigma S^2$ of graphene should be improved. Some experimental efforts have been made in this direction. Wang et al.\cite{WangCR-11PRL} observed enhanced $S$ below room temperature in a dual-gated BLG device, resulting from the opening of a band gap by applying a perpendicular electric field on BLG. Additionally, the Seebeck coefficient and power factor of FLG could be enhanced at high temperature ($>$ 500 K) by molecular attachments\cite{Sim-11JPCC} and oxygen plasma treatment,\cite{XiaoN-11ACSN} attributed to the band gap opening. By constructing the c-axis preferentially oriented nanoscale Sb$_2$Te$_3$ film on monolayer graphene, both $S$ and $\sigma$ were increased, benefiting from a highway for carriers provided by graphene.\cite{Hong-12PCCP} From a practical point of view, Hewitt et al.\cite{Hewitt-12SM} focused on maximizing the power output of FLG/polyvinylidene fluoride composite thin films by considering the absolute temperature, temperature gradient, load resistance, and physical dimensions of films.

\section{Summary and Outlook}
In summary, graphene is one of the best heat conductors in nature. The exceptionally high thermal conductivity appears in graphene caused by a combination of several unique features, specifically, its low dimension, light atomic mass, and strong $sp^2$ covalent bonding. Important isotope effects have been found in graphene, evidencing the weak strength of other scattering processes. Moreover, strong size effects and (quasi-)ballistic thermal transport at room temperature have been observed in graphene due to its long phonon MFP.

We also show possible ways to shorten the phonon MFP, including coupling to a substrate, constructing narrow GNRs with rough boundaries, introducing weakly coupled interfaces, etc. These approaches helps us decrease thermal conductivity of graphene for thermal insulation and thermoelectric applications. We finally discuss the challenges of using graphene for thermoelectrics and possible strategies to overcome the challenges.

Significant progresses have been made in researches on thermal conduction in graphene in the past few years. However, there are still some important fundamental problems unresolved. For example, the change of the scattering strength induced by low dimension has not been well studied. Many quantum effects, like coherent scattering and weak localization, might be important in thermal transport but remain almost unknown (at least in experiments). Whether Fourier's law holds in low-dimensional systems or not is still a controversial issue, and what are the key factors to drive ballistic-diffusive transport is not well understood. On thermoelectrics, the enhancement of $ZT$ requires suppressing thermal conduction while keeping electrical conduction less affected. More works are deserved for designing approaches to effectively decouple electrons and phonons.

We acknowledge the support of the Ministry of Science and Technology of China (Grant Nos. 2011CB921901 and 2011CB606405), and the National Natural Science Foundation of China (Grant No. 11334006).

%\bibliography{myref}
%Merlin.mbs v4.21 2009-07-09.
%

\end{document}